\documentclass[aps,prl,twocolumn,showpacs,amsmath,amssymb]{revtex4}
\usepackage{graphicx}% Include figure files
\usepackage{dcolumn}% Align table columns on decimal point
\usepackage{bm}% bold math
\usepackage{amsmath} 

\begin{document}

\title{Out-coupling vector solitons from a BEC with time-dependent interatomic forces}
\author{ David Feijoo, \'Angel Paredes and Humberto Michinel}
\affiliation{$^1$\'Area de \'Optica, Facultade de Ciencias de Ourense,\\ 
Universidade de Vigo, As Lagoas s/n, Ourense, ES-32004 Spain.
}

\begin{abstract}
%----------------------------   ABSTRACT  ------------------------------------
We discuss the possibility of emitting vector solitons from a two-component
elongated BEC by manipulating in time the inter- or intra-species scattering lengths
with Feshbach resonance tuning. We present different situations which do not
have an analogue in the single species case. In particular, we show vector
soliton out-coupling by tuning the interspecies forces, how the
evolution in one species is controlled by tuning the dynamics of the other, and
how one can implement the so-called supersolitons.
The analysis is performed by numerical simulations of the one-dimensional
Gross-Pitaevskii equation. Simple analytic arguments are also presented in order to
give a qualitative insight.

\end{abstract}
%-----------------------------------------------------------------------------


\maketitle

%--------------------------------------------------------------------------------

\section{I. Introduction}

Since its discovery in 1995 \cite{Anderson95}, the Bose-Einstein condensate (BEC) has been a field which
has attracted a large amount of research effort both from the experimental and
theoretical fronts. Among many interesting research tracks, the design of cold atom
interferometers stands as an exciting possibility \cite{interf}. The external control of 
the condensed atomic clouds is a key factor for any such application. This fact highlights the
importance of the theoretical understanding of atom manipulation protocols and, of course, of
their experimental implementation.

Bright solitons are bunches of energy or matter whose shape does not change upon
 propagation, due to non-linear
effects. They have been discussed in very different contexts such as quantum field theory
or non-linear optics and were first observed in the frame of BECs in \cite{solitons1,solitons2}.
The fact that solitons remain robust and undistorted makes them interesting for the goal
of building precise matter-wave interferometers \cite{parker}-\cite{paper1}. It is well known that they can only be
stable in one-dimensional settings and for that reason we will concentrate in elongated
traps such the dynamics can be effectively reduced to a line \cite{Perezgarcia98}.
In creating such solitary waves, it is essential to have inter-atomic attractive forces. 
The tuning of Feshbach resonances \cite{FB1} --- see \cite{Fesh-review} for a review ---
 is instrumental for this purpose: the scattering length
of atom-atom collisions drastically changes when colliding atoms 
couple resonantly to a molecular bound state, and this
 can be externally controlled by lasers and/or magnetic fields.

The present theoretical contribution introduces protocols of atom manipulation that
can be useful in this framework. Ways of out-coupling self-trapped matter  from a 
shallow potential in a coherent
manner are discussed. There are many previous works following similar strategies, but
we will focus in the possibility of working with time-dependent interactions \cite{Carr04}
in multicomponent BECs,
which allows a high degree of control over the number and properties of the emitted soliton pairs
\cite{paper1}.
In fact, this paper is a natural generalization of \cite{paper1}, extending the 
discussion to
phenomena constrained to multicomponent BECs.
Solitary waves in which different fields are involved in producing the self-trapping 
are called vector solitons. They were first studied by Manakov in the context of non-linear
optical fibers \cite{Manakov}, and have been thoroughly discussed for BECs, see for instance
\cite{xliu,csire,hewang} and references therein. It is important to notice that the Feshbach resonance
tuning technique has been experimentally accomplished in multicomponent condensates
\cite{multiFB}.

In section II, we fix notation and
 present the Gross-Pitaevskii equations (GPEs) used for the analysis.
In sections III-V, we discuss, by numerically solving the
evolution equations, manipulations that allow to out-couple 
solitonic excitations from a shallow trap in which a two-component BEC is initially
confined. The general scheme underlying the different protocols is the following:
first, strong inter-atomic repulsion is induced such that part of atomic cloud is pushed
out of the trap. Then, by tuning again the internal forces, a instability is produced, 
eventually leading to the formation of the desired structures.
We study different situations in which the multicomponent nature of the 
condensate is essential, not having an analogue in the single species case. 
In particular, in section III we show how the process can be carried out by tuning the
{\it interspecies} scattering length. In section IV, we demonstrate how one species can
be indirectly manipulated by tuning the intra-species interaction of the other one.
In section V, we analyse how a so-called super-solitonic excitation 
\cite{supersolitons} can be obtained in the
same framework. 

It is worth noticing  that very interesting phenomena also occur when the two species in the condensate are
different hyperfine states of the same alkali atom. In that case, it is possible to
implement Rabi couplings with an external laser beam in order to transfer atoms
from one species to the other one \cite{zheng}. However, we will not explore this 
possibility in the present work.

%---------------------------------------------------------------------------------
%%%%%%%%%%%%%%%%%%%%%%%%%%%%%%%%%%%%%%%%%%%%%%%%%%%

\section{II. GPEs and initial conditions}
The zero temperature mean-field theory description
of a collection of identical bosons is the Gross-Pitaevskii equation \cite{GP}.
For a multicomponent BEC, there is a coupled system of equations for the evolution
of the atom densities of the different elements:
\begin{equation}
i \,\hbar \frac{\partial \Psi_i}{\partial t}=
-\frac{\hbar^2}{2m_i}\nabla^2 \Psi_i + V_i\,\Psi_i
+ U_{ij} |\Psi_j|^2 \Psi_i
\label{GP1}
\end{equation}
where $m_i$ corresponds to the mass of the $i$'th species, the $V_i$ are
external potentials and the $U_{ij}=2\pi\,\hbar^2a_{ij}/m_{ij}$ parameterize the
inter-atomic forces in terms of the $s$-wave scattering lengths and the reduced masses
$m_{ij}=m_im_j/(m_i+m_j)$. It is worth pointing out that this usually written expression
has been challenged \cite{geltman}. In any case, the precise relation between the
$U$'s and the $a$'s is not essential for our analysis, as long as the inter-atomic forces
can be externally tuned.
The normalization conditions are $\int |\Psi_i|^2 d^3 \vec r = N_i$, where $N_i$ is the
number of atoms of the corresponding species.

In this paper, cigar-shaped atomic traps will be considered: a strong potential
traps the atoms in two directions and the dynamics is reduced to a quasi-one-dimensional
problem. The reason is that it is in this highly asymmetric traps where stable solitons can be
created \cite{Perezgarcia98}.  In the longitudinal dimension ($z$) the system is weakly confined by a shallow optical dipole trap. Thus we take:
\begin{equation}
V_i =  V_{\perp,i} + V_{z,i}(z)=\frac12 m_i \omega_{i\bot}^2(x^2 + y^2) + V_{z,i}(z) 
\end{equation}
where $V_{z,i}(z)$ is defined with a Gaussian shape characterized by a depth potential $V_{0}$ and a proper
width $L$.
In order to dimensionally reduce Eq. (\ref{GP1}), one can write
an approximate solution with separate variables: 
\begin{equation}
\Psi_i = e^{-i\,\omega_{i\bot} t} \sqrt{\frac{m_i \omega_{i\bot}N_1}{\pi\,\hbar\, r_\bot}}
\exp\left(-\frac{m_i \omega_{i\bot}}{2\hbar}(x^2+y^2)\right) \psi_i (z)
\end{equation}
where the length scale $r_\bot= \sqrt{\hbar/m_1 \omega_{1\bot}}$ has been introduced.
Multiplying Eq. (\ref{GP1}) by $\Psi_i^*$, integrating out the $x,y$ dependence 
and defining $\kappa= m_1/m_2$ yields,
for the two-component case:
\begin{eqnarray}
i \frac{\partial \psi_1}{\partial \tau} = - \frac12 
\frac{\partial^2 \psi_1}{\partial \eta^2} + f_1 \psi_1 +
g_{11} |\psi_1|^2 \psi_1 + g_{12} |\psi_2|^2 \psi_1 \nonumber\\
i \frac{\partial \psi_2}{\partial \tau} = - \frac{\kappa}{2} 
\frac{\partial^2 \psi_2}{\partial \eta^2} + f_2 \psi_2 +
g_{12} |\psi_1|^2 \psi_2 + g_{22} |\psi_2|^2 \psi_2
\label{GP}
\end{eqnarray}
A number of dimensionless quantities have been defined:
\begin{eqnarray}
\tau = \omega_{1\bot} t,\qquad \eta =\frac{z}{r_\bot},\qquad
f_i= \frac{V_{z,i}}{\hbar\,\omega_{1\bot}},
\end{eqnarray}
together with:
\begin{eqnarray}
g_{11}=\frac{2N_1 a_{11}}{r_\bot},\qquad 
g_{22}=\frac{2N_1 a_{22}\gamma}{r_\bot }
,\nonumber \\
g_{12}=\frac{2N_1 a_{12}}{r_\bot}\frac{1+\kappa}{1+\kappa/\gamma},\qquad 
\label{gs}
\end{eqnarray}
with $\gamma=\omega_{2\bot}/\omega_{1\bot}$.
The dimensional reduction leading to (\ref{GP}) have been used for instance 
in \cite{kasamatsu}. Slightly different reduction schemes have been employed
in \cite{xliu}, \cite{csire}, \cite{hewang}, yielding still equations (\ref{GP}),
but with certain modifications of the relation between the physical parameters
and the adimensional ones, Eqs. (\ref{gs}).

The new normalization conditions are:
\begin{equation}
\int |\psi_1|^2 d\eta = 1 \,\,,\qquad
\int |\psi_2|^2 d\eta = \frac{N_2}{N_1}
\end{equation}
In the absence of external potential $f_i=0$, equations (\ref{GP}) have
trivial $\eta$-independent stationary solutions. These solutions are perturbatively
unstable under certain conditions and, in particular, their evolution can yield
bunches of self-trapped matter. This modulational instability has been discussed
for two-component GPEs in a series of papers  \cite{kasamatsu}, \cite{vector_modul}, generalizing the 
classical analysis of the single species case \cite{modulational}. It turns out that
stability requires both intraspecies scattering lengths to be positive $g_{11}>0$, $g_{22}>0$
whereas the interspecies interaction has to be sufficiently small $g_{12}^2 < g_{11}g_{22}$.
Thus, instabilities appear for large enough interspecies interaction irrespective of its attractive
or repulsive nature.

In all cases below, Gaussian traps \cite{Stamper98,Martikainen99} will be considered:
\begin{equation}
f_1=f_2={\tilde V_0} \left[1-\exp\left(-\frac{\eta^2}{{\tilde L}^2}\right)\right] \equiv f
\label{trap}
\end{equation}
where $\tilde V_0= V_0/{\hbar \omega_{1\perp}}$ and $\tilde L=L/r_\perp$, so 
atoms can overcome the shallow potential and be out-coupled from the
trap.

The analysis of this paper is based on Eqs. (\ref{GP}). The initial conditions
for the subsequent time evolution will be given by a collection of atoms confined
in the shallow trap. In order to introduce sensible initial conditions, 
we will use a variational
approach within the Thomas-Fermi (TF) approximation, 
which amounts to neglecting the $\eta$-derivative terms in (\ref{GP}).
It is known that the TF approximation yields less accurate results for
the multicomponent BEC than for the single species one \cite{esry}. Nevertheless, since the
stationary ground state is not our main matter of concern and we will only use the
Thomas-Fermi profiles to provide reasonable initial conditions, we will stick to 
it for simplicity (we have checked that the precise form of the initial functions
is not crucial, see section III-B). 
In particular, we assume the initial profiles are Gaussian:
\begin{equation}
\psi_1|_{\tau=0} =\frac{\pi^{-\frac14}}{\sqrt{w_{1,0}}} e^{\frac{-\eta^2}{2w_{1,0}^2}},\quad
\psi_2|_{\tau=0} =\sqrt{\frac{N_2}{N_1}}\frac{\pi^{-\frac14}}{\sqrt{w_{2,0}}} e^{\frac{-\eta^2}{2w_{2,0}^2}}\,
\end{equation}
and determine $w_{1,0}$, $w_{2,0}$ by minimising the functional
$E_{TF}=\int_{-\infty}^\infty {\cal E}_{TF}  d\eta$ where:
\begin{equation}
{\cal E}_{TF}= f\,(|\psi_1|^2 + |\psi_2|^2) + \frac{g_{11}}{2}|\psi_1|^4+
\frac{g_{22}}{2}|\psi_2|^4+g_{12}|\psi_1|^2|\psi_2|^2
\label{etf}
\end{equation}
In the simplest, symmetric case $N_1=N_2$, $g_{11}=g_{22}\equiv g = -g_{12}+\tilde L\sqrt{\pi}\tilde V_0$, one 
finds $w_{1,0}=w_{2,0}=\tilde L$. Evidently, the $g_{ij}$ entering Eq. (\ref{etf}) are the values before the beginning of the manipulations, {\it i.e}. in the ground state of the trapped atoms.
%{\it before} $\tau =0$.

The regime of validity of the GPE and its 1-D reduction has been 
widely discussed in the literature,
see for instance \cite{Carr04,carrbrand2} and references therein. 
The dynamics is confined to one dimension if the 
interaction energy is not enough to excite higher modes in the
transverse harmonic oscillator. 
In terms of adimensional quantities, this means, roughly
$\sum_j |g_{ij}||\psi_j|^2 < 1 $ for $i=1,2$ 
In all examples below, we will require this
condition to hold at the center of the trap for the initial profiles.
Thermal corrections 
of the GPE, which can eventually
affect the evolution of the atom densities \cite{sinha} or the coherence properties of the
wavefunctions \cite{petrov}, will be disregarded in the analysis. 
It is also worth mentioning that large scattering lengths
come together with high 3-body recombination rates which result in the depletion of
the condensate --- this, in fact, is the basis of a usual procedure for experimentally
determining Feshbach resonances. However, the protocols that will be discussed in the
following do not need extreme values of $a_{ij}$ and we will not take this phenomenon
into account.

% As in the case of only one species, the one-dimensional
%(1D) approximation remains sensible as long as
%the atoms are not energetic enough to get excited and probe
%the transverse directions. The consequence is that a brief discussion about the
%validity and limitations of Eq. (\ref{GP}) is necessary again, so as to know the grade 
%of feasibility of the approximation leading to reduce
%Eq. (\ref{GP1}) to its 1D counterpart Eq. (\ref{GP}). It is important to say that, with attractive
%interactions, collapses not describable with (\ref{GP}) may occur.
%Avoiding collapse in the transverse 2D space or collapse of a
%single soliton requires conditions that can be taken using approximately the one species case \cite{paper1},
%so $-(g_{11}+g_{12})|\psi|^2 < 0.93$ and $-(g_{11}+g_{12})N_{ais} < 1.25$ , where $N_{ais}$ stands
%for the number of atoms in a soliton. 

In the following, with the aim of keeping some generality, the dimensionless notation is used in most of the equations. Figures will 
 present examples
with dimensionful  values of the parameters. The translation between both formalisms is straightforward
using the definitions given in this section.  For instance, a typical value for $\omega_{1\bot}$ may be $10^3$s$^{-1}$.
Then, if one considers $^{7}$Li atoms, $r_\bot \approx 3\mu m$. With this in mind, one can get a 
ballpark estimate of the physical meaning of the dimensionless parameters:
 the unit of $\tau$ would correspond to 1 ms and the unit of $\eta$ to $3\mu m$. 
 Many examples are provided by choosing the different scattering lengths in order to have figures
 which illustrate neatly  the general ideas. Notwithstanding, sections III-C, IV-B, V-A introduce 
 cases using values of the scattering lengths and magnetic Feshbach resonances reported in the
 experimental literature for concrete atomic mixtures.

%However, when eventually the different protocols are experimentally realized, one should
%recheck the validity for the corresponding dimensionful specifications of that particular case.

%We close this section by mentioning that
%The plots in the following will be displayed with these dimensionful parameters.
%Needless to say, for different masses or trapping frequencies, the translation to
%different dimensionful quantities is straightforward.

%%%%%%%%%%%%%%%%%%%%%%%%%%%%%%%%%%%%%%%%%%%%%%%%%%%

\section{III. Soliton emission with a time-dependent interspecies scattering length}

The most obvious phenomenon that is present in multicomponent BECs but not
in single species ones is that of inter-species forces. We will thus start our
analysis by showing how by tuning this parameter it is possible to out-couple
vector bright-bright solitons from a sample initially trapped in a shallow potential.
Due to the large space of parameters there is a plethora of possibilities but, in order
to focus in the desired dynamics, we will consider that only $g_{12}$ varies in time.

As the simplest example, let us consider $\kappa=1$,
$N_1=N_2$, $g_{11}=g_{22} \equiv g$ and, for $\tau<0$ let us take
$g_{12}=0$. Thus, initially, the atom clouds for both species are identical
and are well approximated by the Thomas-Fermi profile:
\begin{equation}
\psi_1 = \psi_2 = \frac{1}{\pi^\frac14 \sqrt{\tilde L}}\exp\left(-\frac{\eta^2}{2{\tilde L}^2}\right)
\end{equation}
if we assume that $g=\sqrt{\pi} \tilde V_0 \tilde L$.
Notice that an initial relative phase would play no r\^ole in the evolution.
By turning on $g_{12}= \tilde g_{12}>0$ in the time interval $0<\tau<\tau_s$, the cloud expands
out of the trap due to the repulsive forces. Then, at a given time $\tau_s$, one can tune
$g_{12} = \hat g_{12}< - g < 0$ in order to spark modulational instability and vector
soliton formation. The value of $\tau_s$ has to be chosen properly. If it is too small,
repulsion would not be enough for the solitons to leave the trap. If it is too large,
the cloud gets to disperse before tuning to $g_{12}<0$ and soliton formation is impeded.
See section III-A for a related discussion.

Due to the symmetry between both species in this 
simplified case, it is clear that the condition $\psi_1 = \psi_2$ can be a
solution for the whole evolution and thus, formally, the system of equations 
is reduced to that of \cite{paper1}. One could wonder whether this $\psi_1 = \psi_2$
solution is stable. In fact, that is the case for the soliton
emission process. We will come back to this question in section III-B.
Adapting the results of \cite{paper1}, 
 a rough estimate to how the number of vector solitons emitted varies with
the different quantities can be given:
\begin{equation}
N_s \approx 2\left[ \frac{c_1\sqrt{-(g+\hat g_{12})}(c_2 \sqrt{\tilde g_{12}+g} -\tilde L)}{(\tilde g_{12}+g)^\frac14}
\right]
\label{nsest}
\end{equation}
where $c_1$, $c_2$ can be fitted to data.
The term $(c_2 \sqrt{\tilde g_{12}+g} -\tilde L)$ 
becomes negative
for small $g_{ij}$, what might seem surprising. The point is that,
since the solitons feel the Gaussian trap, they need a sufficient initial boost to be able to escape from it.
Distinctly, the boost is provided by the repulsive interactions in the expansion period. 
 If this repulsion is not large enough, the estimated
$N_s$ becomes negative, meaning that solitons, even if produced, are not able to 
overcome the trapping potential.

\begin{figure}[htb]
\includegraphics[width=0.48\textwidth]{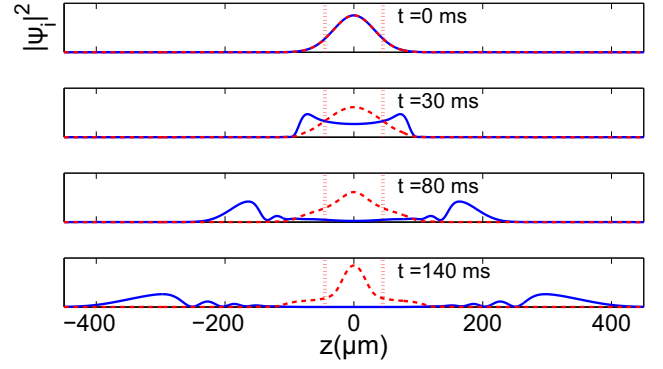}
\caption{(Color online). An example of evolution in time of the
atom densities of both species $|\psi_1|^2(z)$ (solid blue line) for lithium,
$|\psi_2|^2(z)$ (red dashed line) for sodium.
These simulation have been performed with parameters $a_{11}=a_{22}=0.3 nm$,
$N_1=N_2=50000$, $V_0=\hbar \omega_\perp/15$, $L=45 \mu m$. The inter-species scattering length parameter is 
tuned from $a_{12}=0$ to
$\tilde a_{12}=0.45 nm$. The red dashed vertical lines located at
$z=\pm {L}$ indicate the size of the shallow trap.
}
\label{fig0}
\end{figure}

Let us now turn to a non-symmetric case $\kappa \neq 1$. 
The Thomas-Fermi initial profiles will be the same as before, but the subsequent
evolution differs since it cannot be reduced to a single second order equation any more.
Before showing the results of some characteristic simulations, we provide some analytic
insight of the problem. The first step of the process is the expansion of the cloud due
to repulsive $\tilde g_{12}$. A simple modelization can be constructed in the 
averaged Lagrangian formalism \cite{AL}. A variational ansatz is considered:
\begin{equation}
\psi_j =A_j(\tau) e^{\frac{-\eta^2}{2w_j(\tau)^2}}e^{i(\mu_j(\tau)+\eta^2 \beta_j(\tau))}
\end{equation}
Notice that the ansatz for the initial conditions was taken consistent with this form.
These expressions are inserted
into $\bar {\cal L}=\int_{-\infty}^\infty {\cal L} d\eta$ where:
\begin{equation}
{\cal  L} = \sum_{j=1}^2 \left[
-\frac{i}{2}(\psi_j \partial_\tau \psi_j^* -\psi_j^* \partial_\tau \psi_j)
+ \frac12 |\partial_\eta \psi_j|^2 + {\cal E}_{TF}
\right]
\end{equation}
where ${\cal E}_{TF}$ can be found in Eq. (\ref{etf}).
From the averaged Lagrangian $\bar {\cal L}$ one can write the Euler-Lagrange equations for the eight
real fields $A_j(\tau),w_j(\tau),\mu_j(\tau),\beta_j(\tau)$. Since the main interest is to 
compute the evolution of the sizes of the clouds $w_j(\tau)$, we just write down the relevant
equations:
\begin{eqnarray}
\ddot w_1(\tau)= \frac{1}{w_1(\tau)^3}
- 
 \frac{2 \tilde L\,\tilde V_0 w_1(\tau)}{({\tilde L}^2+w_1(\tau)^2)^\frac32}
 + 
 \nonumber \\
+ \frac{g_{11}}{\sqrt{2\pi}\,w_1(\tau)^2}
 + \frac{N_2^2}{N_1^2}\frac{2g_{12}w_1(\tau) }{\sqrt{\pi}(w_1(\tau)^2+w_2(\tau)^2)^\frac32} \,. \\
\ddot w_2(\tau)= \frac{\kappa^2}{w_2(\tau)^3}- 
 \frac{2\kappa\,\tilde L\,\tilde V_0 w_2(\tau)}{({\tilde L}^2+w_2(\tau)^2)^\frac32}
 + 
 \nonumber \\
+ \frac{N_2^2}{N_1^2}\frac{g_{22}}{\sqrt{2\pi}\,w_2(\tau)^2}
 + \frac{2\kappa\,g_{12}w_2(\tau) }{\sqrt{\pi}(w_1(\tau)^2+w_2(\tau)^2)^\frac32} \,.
\end{eqnarray}
This suggests that, if the expansion is driven by the interspecies repulsion $g_{12}$,
and assuming $N_1 \approx N_2$,
the less massive cloud of atoms expands faster.
The simulation of figure \ref{fig0}
 confirms this expectation. Taking $\kappa=0.3061$, as it corresponds to a 
 lithium-sodium mixture, the simulation was performed 
 with the dimensionless parameters  $g_{11}=g_{22}=1$,
$N_1=N_2=50000$, $\tilde V_0=1/15$, $\tilde L=15$ and
$\tilde  g_{12}=15$. These 
quantities have been introduced fixing $\omega_{\bot}=\omega_{1\bot,2\bot}=10^3$s$^{-1}$, 
and the rest of dimensional parameters presented in the caption.  
The figure also shows how bumps appear in
the wave-functions even in the absence of attractive interaction. Notice these bumps do
not have a solitonic character and tend to spread out.

\begin{figure}[htb]
\centering \resizebox*{1\columnwidth}{0.6\columnwidth}{\includegraphics{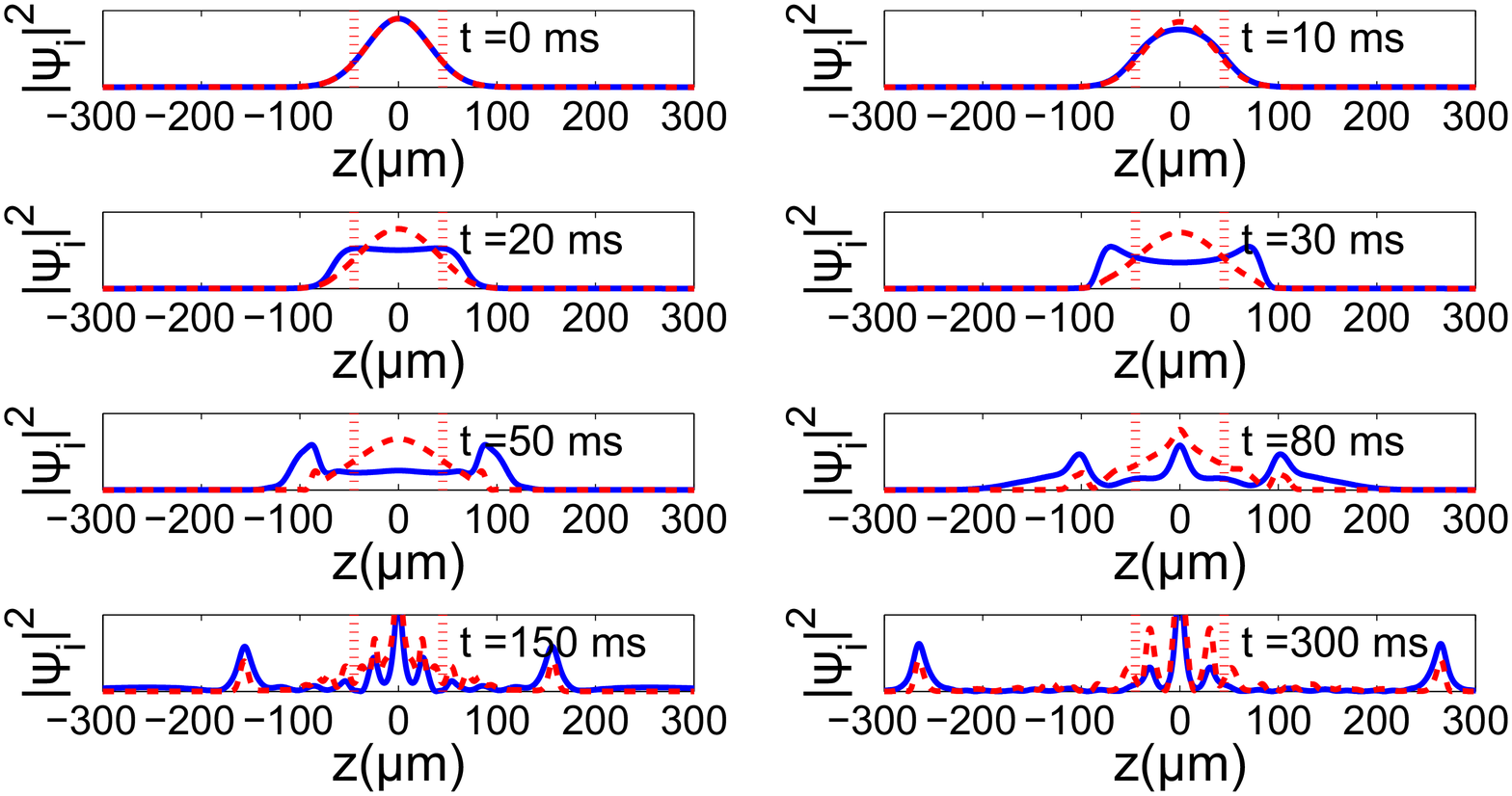}}
\caption{(Color online). Simulation with parameters $\kappa=0.3061$, $a_{11}=a_{22}=0.3 nm$,
$N_1=N_2=50000$, $V_0=\hbar \omega_\perp/15$, $L=45 \mu m$. Taking $t_s=25 ms$, the inter-species scattering length parameter is tuned 
from $a_{12}=0$ for $t<0$ to
$\tilde a_{12}=0.45 nm$ for $0<t<t_s$ to $\hat a_{12}=-0.18 nm$ for $t>t_s$. The solid blue line corresponds to lithium and the dashed red line to sodium.
}
\label{fig1}
\end{figure}

The second step of the process is soliton formation after tuning $g_{12}$ to a negative
value therefore sparking modulational instability. 
We show the result of numerical simulations in figure \ref{fig1}. As a representative example we
have taken again a lithium-sodium mixture, that is $\kappa=0.3061$,  and the dimensionless parameters $g_{11}=g_{22}=1$,
$N_1=N_2=50000$, $\tilde V_0=1/15$, $\tilde L=15$, $\tau_s=25$, $\tilde g_{12}=15$ before $\tau_s$ and $\hat
 g_{12}=-6$ after $\tau_s$. These quantities were introduced fixing $\omega_{\bot}=\omega_{1\bot, 
 2\bot}=10^3$s$^{-1}$ and their correspondent dimensional parameters are presented in the caption of the figure. It 
 can be observed how
the repulsive interspecies force pushes part of the atomic cloud out of the trap,
with the lighter atoms in front. The sudden
change of $g_{12}$ leads to the grouping of a bunch of atoms.
Even if the process is not symmetric under the change $\psi_1 \leftrightarrow \psi_2$, the attractive
interactions eventually groups a fraction of the atoms into vector solitons which escape from the trap thanks to
their own inertia. In this case, a single soliton pair is produced.

\begin{figure}[htb]
\includegraphics[width=0.47\textwidth]{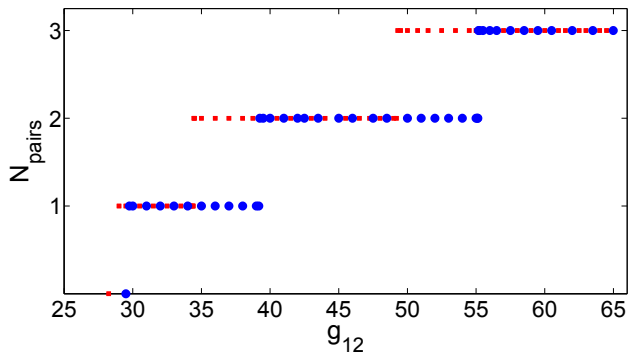}
\caption{(Color online). Number of out-coupled vector soliton pairs vs. the
inter-species scattering length driving the expansion $\tilde g_{12}$. Blue circles correspond to $\kappa = 0.31$ (a lithium-sodium mixture) and red squares to $\kappa = 0.47$ (a potassium-rubidium mixture).
The rest of parameters are fixed as in figure \ref{fig1}. The conversion to $a_{12}$ can be made using Eq. (\ref{gs}), taking $\omega_{\bot}=10^3$s$^{-1}$ and $r_\bot \approx 3\mu m$ (for the lithium-sodium mixture) or $r_\bot \approx 1.25 \mu m$ (for the potassium-rubidium mixture). 
}
\label{fig2}
\end{figure}

In the asymmetric case $\kappa \neq 1$, it does not seem possible
to provide a simple reliable formula which estimates 
the number of emitted solitons, similar
to Eq. (\ref{nsest}). However, some qualitative features remain:
below a certain value of $\tilde g_{12}$, the repulsive force is not
strong enough and no vector soliton comes out. Then, the number of
solitons grows with increasing $\tilde g_{12}$ since more atoms are out-coupled from
the trap. These features can be easily appreciated
in figure \ref{fig2}, which shows the results of a sample of simulations.
The figure also shows that the number of emitted solitons slightly decreases 
with $\kappa$.
The dependence on $\hat g_{12}$ is weaker: as long as this attraction is large enough to
pack the out-coming atoms into vector solitons, it hardly affects its number.

\subsection{A. Possible limitations in the time of tuning}

The method of soliton emission presented in this paper has
considered, until now, instant modifications in the inter-species scattering length. This is
 an ideal assumption, since such a sudden tuning 
from zero to the chosen value is impossible. Consequently, it is important to analyse
 what happens when this parameter is changed in a smoother way. In the following,
  different simulations are presented in order to check the possible time
   limitations in the $a_{12}$ tuning to get the right initial triggering
and soliton outcoupling processes. The parameters will be those displayed in figure \ref{fig1}
  with the only difference that $a_{12}$ is now defined by: %In order to get a right initial triggering and soliton outcoupling, it is important to analyse how fast the change in the inter-species scattering length must be. 

 \begin {equation} a_{12}=\begin{cases} 0 &\mbox{if } t\leq 0 \\
\frac{\tilde a_{12} }{t_1} t &\mbox{if } 0< t< t_1 \\
\tilde a_{12} &\mbox{if } t_1\leq t\leq t_2 \\
\frac{\hat a_{12}(t-t_2)-\tilde a_{12}(t-t_3) }{t_3-t_2}  &\mbox{if } t_2<t<t_3 \\
\hat a_{12} &\mbox{if } t\geq t_3 \\
\end{cases} \end {equation}
\\
where the different time instants $t_1$, $t_2$ and $t_3$ were introduced in order to modify $a_{12}$ gradually until reaching its maximum and minimum values $\tilde a_{12}$ and $\hat a_{12}$.
The quantities $\Delta t_1=t_1-0=t_1$ and $\Delta t_2=t_3-t_2$ represent respectively how fast 
each tuning is and they are the subject of study.

\begin{figure}[htb]
 {\centering
  \resizebox*{1\columnwidth}{0.6\columnwidth}{\includegraphics{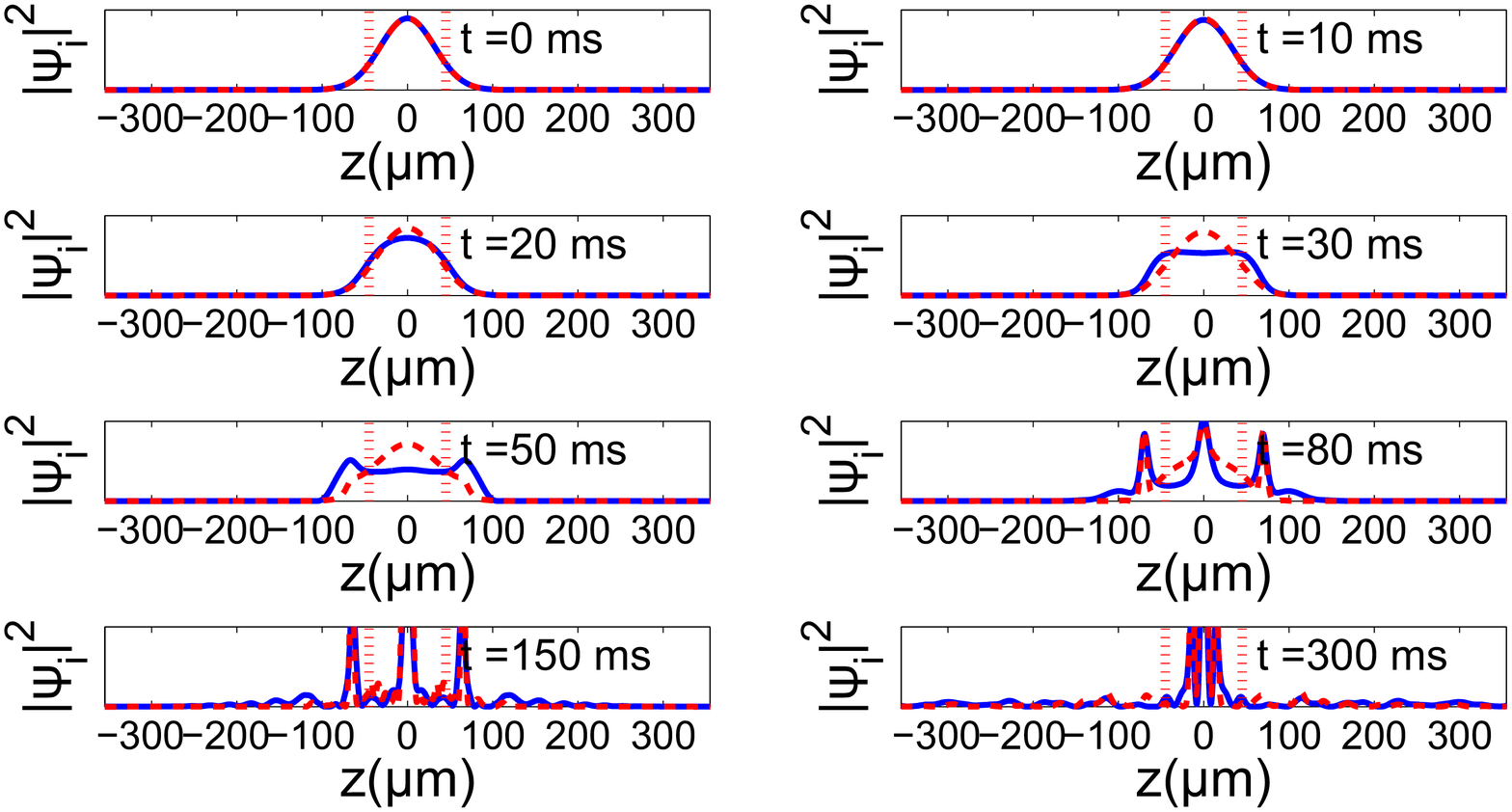}} }
 \caption{(Color online). Simulation with the same parameters that in figure \ref{fig1} except for $a_{12}$.  
 The following values have been taken:
  $t_1=19 ms$, $t_2=25 ms$ and $t_3=27 ms$, so $\Delta t_1=19 ms$ and $\Delta t_2=2 ms$. 
 The solid blue line corresponds to lithium and the dashed red line to sodium.}
 \label{estimation2}
 \end{figure}

 Let us start with the analysis of the tuning for the expansion, that is in $0\leq t \leq t_2$.
 It turns out that the important factor is the kinetic energy acquired by the atom cloud during its
 expansion. 
 If the expansion rate is not enough, solitons  collapse re-entering the trap,
 see Fig. \ref{estimation2}.
 
\begin{figure}[htb]
{\centering 
 \resizebox*{1.0\columnwidth}{0.58\columnwidth}{\includegraphics{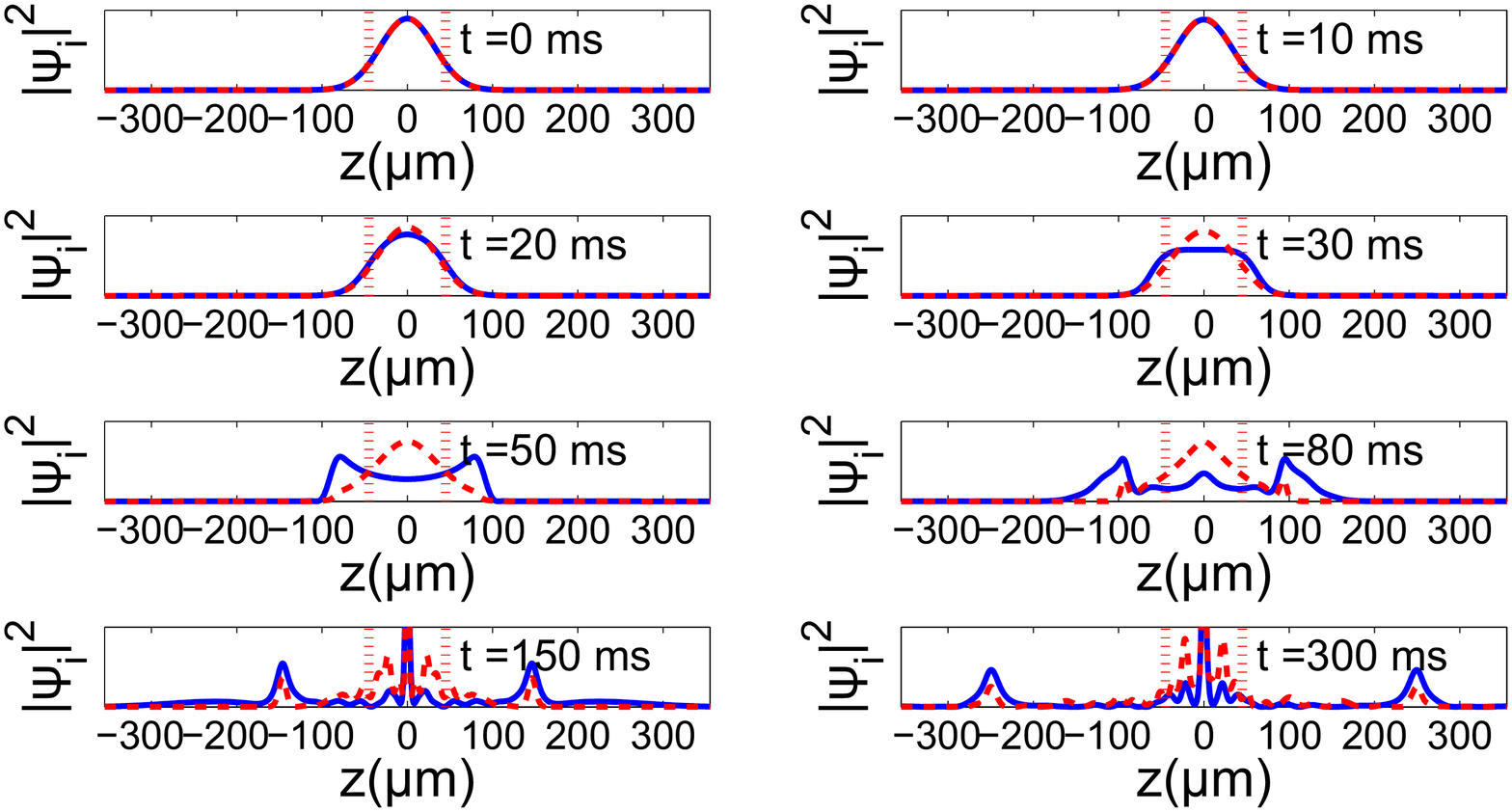}}}
\caption{(Color online). Simulation with parameters $t_1=26 ms$ --- so $\Delta t_1=26 ms$ --- and $t_2=38 ms$,
 $t_3=40 ms$ --- so $\Delta t_2=2 ms$. The rest of parameters are the same that in figure \ref{fig1}. 
  The solid blue line corresponds to lithium and the dashed red line to sodium.}
\label{estimation1}
\end{figure}

 The conclusion is that the time $\Delta t_1$ is not decisive as it can be always compensated
 by delaying $t_2$, see Fig. \ref{estimation1} where soliton emission is similar to that in Fig. \ref{fig1}. 

\begin{figure}[htb]
{\centering
 \resizebox*{1.0\columnwidth}{0.6\columnwidth}{\includegraphics{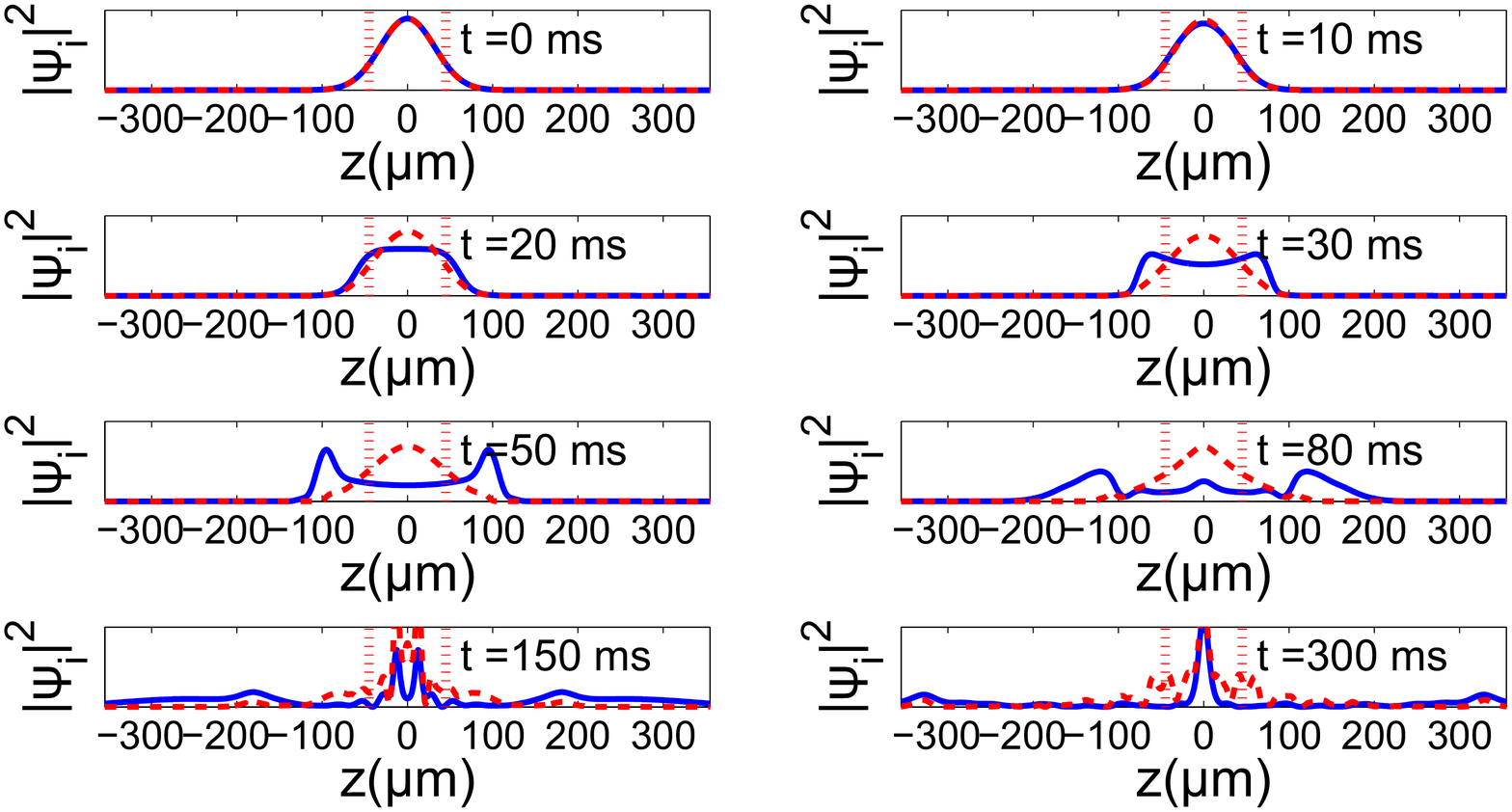}}\par}
\caption{(Color online). Simulation with the same parameters that in figure \ref{fig1} except for $a_{12}$. In
these plots,
 $t_1=5 ms$, $t_2=25 ms$ and $t_3=45 ms$, so $\Delta t_1=5 ms$ and $\Delta t_2=20 ms$. The solid blue line corresponds to lithium and the dashed red line to sodium.}
\label{estimation3}
\end{figure}

Let us turn to the tuning involved in the soliton formation, that is in $t>t_2$.
Our simulations show that there is  a limiting value for $\Delta t_2$ above
which solitons are not produced. The reason is that the atom cloud spreads out excessively
before the self-trapped atom clusters are actually formed.
Fig. \ref{estimation3} provides an example of 
this fact.

The specific constraint on $\Delta t_2$ depends non-trivially
 on all the rest of parameters, but, roughly, a few miliseconds seem enough to 
 get success in soliton formation.

\subsection{B. Introduction of noise in the process}

\begin{figure}[htb]
{\centering \resizebox*{0.9\columnwidth}{0.5\columnwidth}{\includegraphics{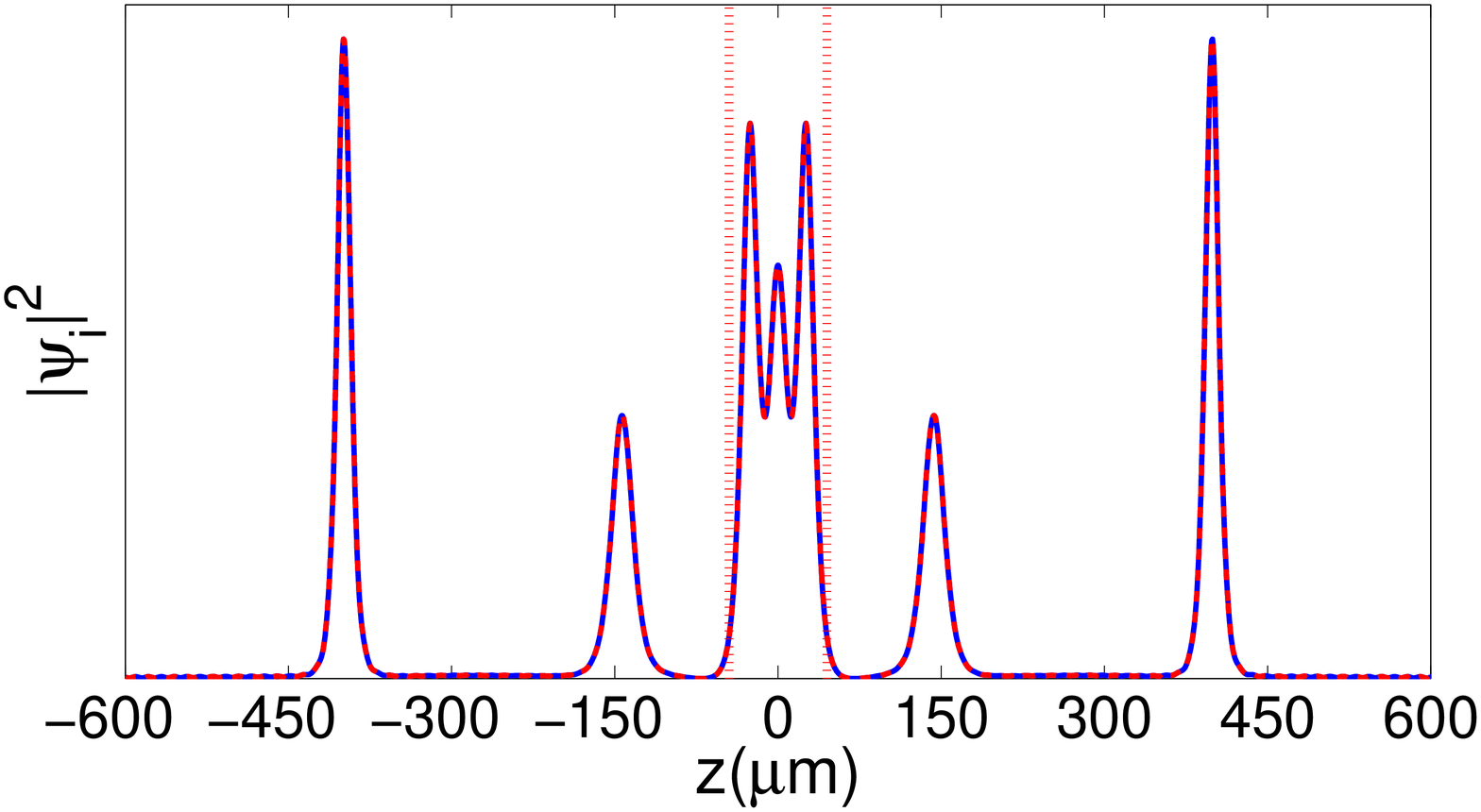}}
\resizebox*{0.9\columnwidth}{0.5\columnwidth}{\includegraphics{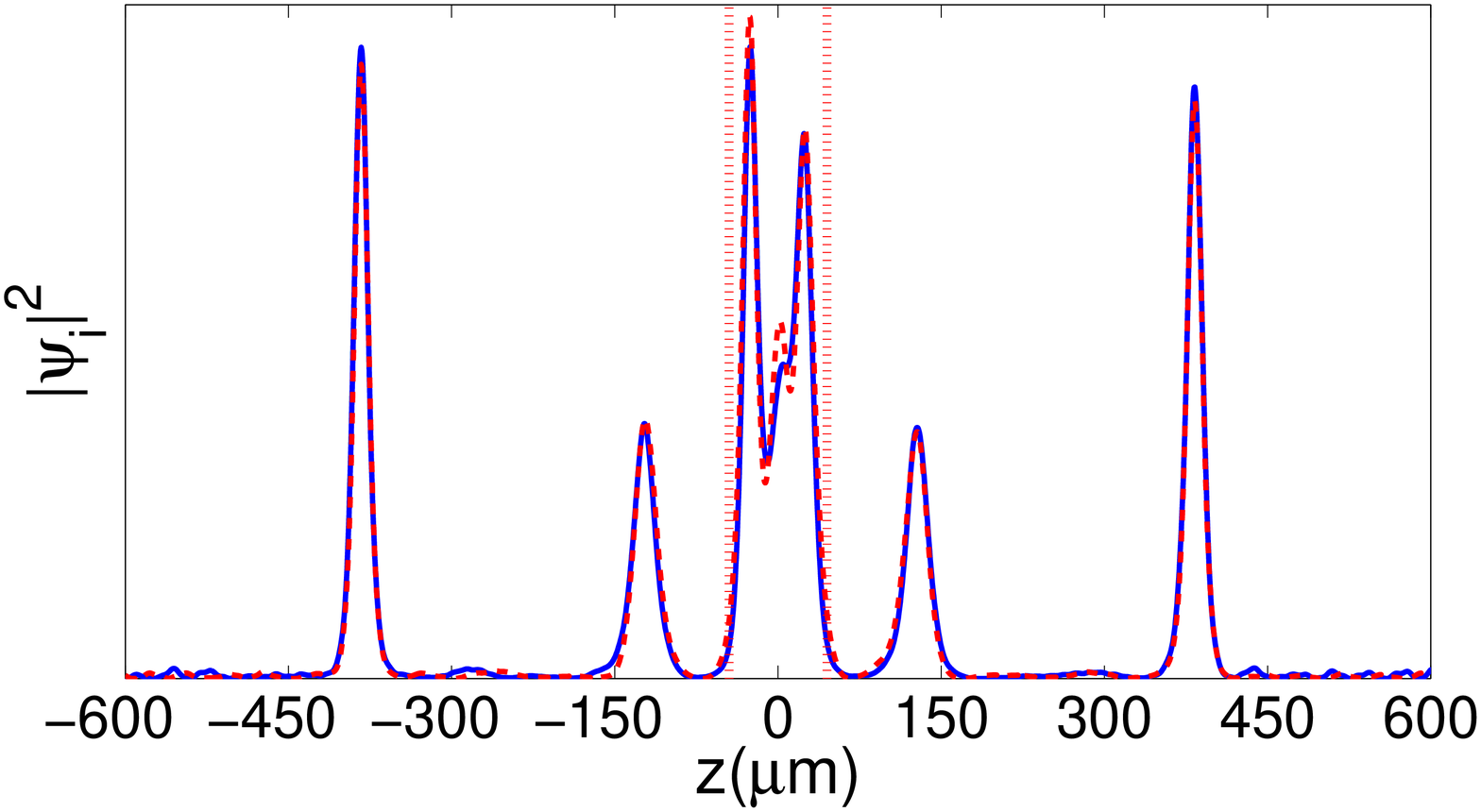}}\par}
\caption{(Color online). Emission of a vector soliton pair in the symmetric case.
All parameters are as in figure \ref{fig1} except for $\kappa=1$.
Above, a noise-less simulation where the result has strictly $\psi_1=\psi_2$.
Below, random noise was introduced in the initial conditions.
}
\label{fig:noise}
\end{figure}

Another relevant question is how robust the results described are. For instance, in the
symmetric $\kappa=1$, $N_1=N_2$ case, it is not obvious whether the evolution with $\psi_1 = \psi_2$
locked to each other is stable. Indeed, in a situation without potential and flat atomic
distributions, the system is modulationally unstable even if all interactions are repulsive
as long as $g_{12}^2 > g_{11}g_{22}$. However, it can be checked with simulations that, in the
processes described, the outcome is well described by the $\psi_1 = \psi_2$ solution. What 
happens is that when the solitons are formed, the differences that had appeared during the expansion
of the cloud are partially washed out. As an illustration, we depict in figure \ref{fig:noise} 
a comparison of the result of simulations with and without random noise introduced in the
initial conditions. The noise was introduced by multiplying the discrete Fourier transform 
of the initial Gaussian profiles by real pseudo-random
numbers extracted from a normal distribution with mean 1 and standard deviation 0.2. 
After this random modification, the wave-functions are normalized by dividing by a global factor.
It can be appreciated how the out-coupled vector soliton pair is almost unaffected by the 
noise.

With this kind of simulations, it can also be checked that the precise form of the initial
profile is not a critical factor for the qualitative evolution of the atom cloud. For this
reason, using Thomas-Fermi in order to insert the initial conditions is a justified approximation.

\subsection{C. Experimental considerations}

Even if the rich structure of Feshbach resonances in different mixtures of
alkali atoms results in an ample set of possibilities for manipulation, it is
clear that not all protocols that one may imagine can be realized in the laboratory.
Around a magnetically tuned Feshbach resonance, the scattering length varies 
as \cite{Fesh-review}:
\begin{equation}
a(B)=a_{bg} \left(1-\frac{\Delta}{B-B_0}\right)
\label{aofB}
\end{equation}
where $B_0$ is the position of the resonance, $\Delta$ its width and $a_{bg}$ the off-resonance
value of the $s$-wave scattering length.  In a two-component mixture, similar expressions hold
for each intra-species interaction and for the inter-species one. Thus, the three relevant scattering
lengths depend on a single externally tunable parameter $B$.

The implementation of the process described in this section requires the possibility of widely tuning
$a_{12}$ while $a_{11}$, $a_{22}$ remain positive and essentially constant. Understanding which are all
the mixtures in which this is possible lies beyond the scope of the present work, but we now argue that
in ${}^7$Li-{}$^{23}$Na it can be accomplished, at least in principle.

For sodium, the background value away from a few narrow resonances is $a_{22}\approx 63 a_0$
where $a_0 \approx 0.053 nm$ is Bohr radius --- see \cite{Fesh-review} and references therein.
${}^7$Li in its $|F=1\ m_f=1\rangle$ hyperfine state
presents a broad resonance with $B_0 \approx 736.8$G, $\Delta \approx -192.3$G and
$a_{bg} \approx -25 a_0$ \cite{solitons1,7Li}. It is worth mentioning that this was the
isotope used in the first experiments showing the formation of bright (single-species)
solitons \cite{solitons1,solitons2}. In the region between the zero-crossing point 
$B_{zc}\approx 544$G and the resonance, the scattering length remains positive and varies
according to Eq. (\ref{aofB}), as experimentally studied in detail in \cite{7Li}.
Several inter-species ${}^7$Li-${}^{23}$Na resonances were predicted in \cite{schuster}
fitting experimental observations for ${}^6$Li-${}^{23}$Na to a theoretical 
coupled-channel model. In particular, there are two of them around
$B_0 \approx 600$G and $B_0 \approx 650$G.  We conclude that by tuning the magnetic
field around these values,
$a_{12}$ can be drastically shifted while $a_{22}$ does not change and $a_{11}$ varies
only mildly, providing a possible experimental scenario in which to realize the suggested
out-coupling of vector solitons.

\begin{figure}[htb]
\centering \resizebox*{1\columnwidth}{0.6\columnwidth}{\includegraphics{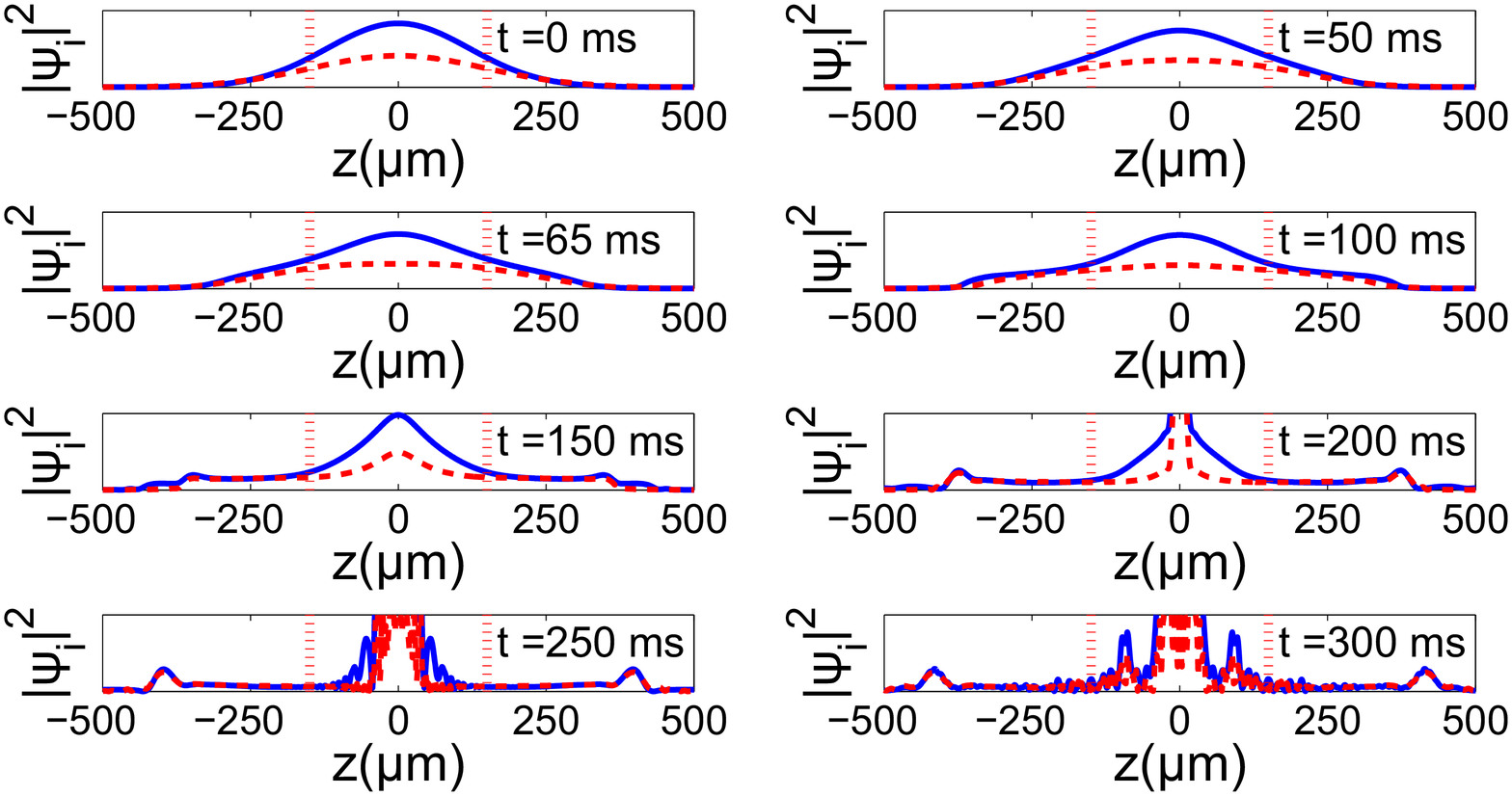}}
\caption{(Color online). Simulation with parameters $\kappa=0.3061$, $a_{11}=30 a_0$, $a_{22}=63 a_0$,
$N_1=50000$, $N_2=30000$, $V_0=0.5 \hbar \omega_\perp$, $L=150 \mu m$. Taking $t_s=65 ms$, the inter-species scattering length parameter is tuned 
from $a_{12}=0$ for $t<0$ to
$\tilde a_{12}=3.3 nm$ for $0<t<t_s$ to $\hat a_{12}=-2.25 nm$ for $t>t_s$. The solid blue line corresponds to lithium and the dashed red line to sodium.
}
\label{figLi-Na}
\end{figure}

Figure \ref{figLi-Na} shows the process of soliton emission for this case. The widths of the initial Gaussian profiles $w_{1,0}$ and $w_{2,0}$ were calculated for this simulation minimising Eq. (\ref{etf}).
It can be observed that the outcome is qualitatively simliar to Fig. \ref{fig1} despite notorious
differences in some of the parameters entering the computation.

%%%%%%%%%%%%%%%%%%%%%%%%%%%%%%%%%%%%%%%%%%%%%%%%%%%

\section{IV. Extracting matter with matter}

In this section we will focus on the possibility of using one species in order to
control the atom cloud of the second one through the inter-species force. This indirect 
manipulation can be specially useful when it is simpler to manipulate one of the 
components of the mixture. We will show that in this fashion it is possible to
out-couple trapped matter waves if the inter-atomic attraction is strong enough.
Notice that the idea is similar to the one underlying the technique that allowed
to create the first mixed BECs \cite{sympa}, where one of the species was cooled down whereas
the temperature of the second one only decreased by interaction with the first.

The simplest way to realize control of the species 2 via species 1 is to consider
$N_1 \gg N_2$. In the limit $N_2 / N_1 \to 0$, the computation can be split in
a two step process. First, one has to solve a non-linear equation for $\psi_1$ in which
the second species can be ignored.  Then, assuming that the
self-interaction of the second species is negligible,
 one can solve the {\it linear}
Schr\"odinger equation for the second species in which the total potential is given
by the sum of the external potential and the effective, time dependent, potential induced 
by the atom cloud of the first species, namely:
\begin{equation}
f_{2,tot} = f_2 + g_{12} |\psi_1|^2
\label{f2tot}
\end{equation}
On the other hand, when
the second species back-reacts on the first
one, the coupled system of equations must be solved.
Two different situations in which these general ideas
may be realized will be considered in turn.

\subsection{A. A soliton extracts matter from a trap}

Let us consider species 2 as starting in the ground state of the linear
Schr\"odinger equation with trapping potential
(\ref{trap}). The values $\tilde {V_0}=1/15$, $\tilde {L}=15$ will be fixed.
The ground state wave-function can be easily computed by a
numerical shooting 
method.
We then compute its evolution when a bright soliton of species 1 traverses the atom cloud
with velocity $v$. 
We take $g_{11},g_{12} < 0$ and:
\begin{equation}
|\psi_1|^2 = - \frac{g_{11}}{4}
\frac{1}{\cosh^2(-\frac{g_{11}}{2}(\eta-(\eta_0+v\,\tau)))}
\end{equation}
By inserting this expression into (\ref{f2tot}), (\ref{GP}), it is possible
to study what happens to species 2 for different values of $g_{11}$, $g_{12}$, $v$.
In order to understand the results, we start by writing down the following
Schr\"odinger equation:
\begin{equation}
-\frac12 \frac{\partial^2 \psi_2}{\partial x^2}
- \frac{s(s+1)}{2\,\cosh^2 x}\psi_2 = E \,\psi_2
\label{eqsolit}
\end{equation}
which, after some redefinitions, including:
\begin{equation}
\frac{g_{12}}{g_{11}} \equiv \frac{s(s+1)}{2}
\end{equation}
can be found to be the equation governing the 
second species in the presence of a static soliton of the
first one. In fact, the $s=1$ case, corresponding to 
$g_{11}=g_{12}$ is called the one-soliton potential, see
for instance \cite{one-soliton}. For $s>0$, the discrete 
spectrum consists of $\lceil s \rceil$ eigenstates 
of energies $-(s-n)^2/2$ with $n=0,1,\dots, \lceil s \rceil -1$,
see \cite{landau}. The continuous spectrum of (\ref{eqsolit}) is
remarkable since the potential proves to be reflection-less
if $s$ is an integer \cite{reflectionless}.

\begin{figure}[htb]
\includegraphics[width=0.48\textwidth]{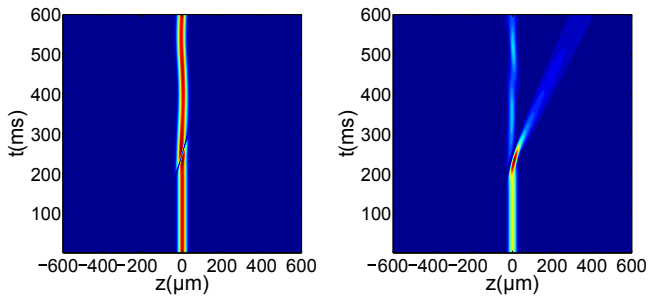}
\caption{(Color online). Contour plots of the time evolution of the wave-function
$|\psi_2|^2$. In both cases $z_0=-150\mu m$, $v=0.6 \mu m/ms$. On the left, $a_{11}=-0.6 nm$, $a_{12}=-1.8 nm$.
On the right,  $a_{11}=-0.6 nm$, $a_{12}=-1.2 nm$.
}
\label{fig:reflect}
\end{figure}

This last observation is important for the problem at hand since, for
$|g_{11}| \gtrsim 1$, it can be observed that if
$\frac{g_{12}}{g_{11}}=1,3,6,10,\dots$,
the soliton passes through
the trap almost without affecting it, so the soliton continuous its path and it does not practically extract atoms of the cloud. %, see the left plot in Fig. \ref{fig:reflect}.
On the other hand, for other values of $\frac{g_{12}}{g_{11}}$ 
--- non-integer $s$ ---, the potential in (\ref{eqsolit}) ceases to be
reflection-less and, indeed, the passing soliton pushes atoms out of the trap. Both  behaviours can be discerned in Fig. \ref{fig:reflect}. The simulations where performed with the dimensionless parameters $\eta_0=-50$ and $v=0.2$ for the two plots, taking $g_{11}=-2$, $g_{12}=-6$ on the left plot and $g_{11}=-2$, $g_{12}=-6$ on the right one. Considering a number of 5000 atoms per soliton and atoms of $^{7}$Li, so that $r_\bot \approx 3\mu$m, the correspondent dimensionful parameters are presented in the caption of the figure.    %,
%right plot in Fig. \ref{fig:reflect}. 
On the right plot, the atoms of the second species exit the
trap in front of the soliton, as a tennis ball hit by a racket, so these atoms are not a solitonic wave. For fixed ${g_{12}}/{g_{11}}$ and $v$,
the number of out-coupled atoms grows with  $|g_{12}|,|g_{11}|$ since the
interaction is stronger.

\begin{figure}[htb]
\includegraphics[width=0.48\textwidth]{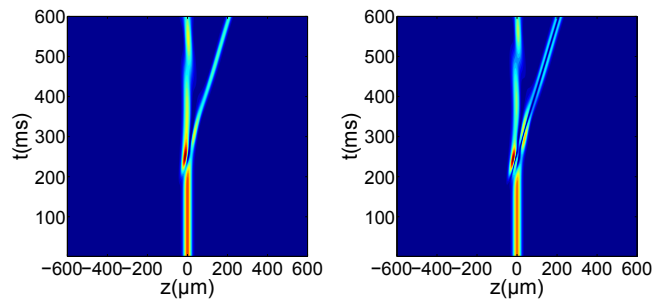}
\caption{(Color online). Contour plots of the time evolution of the wave-function
$|\psi_2|^2$. In both cases $z_0=-150\mu m$, $v=0.6 \mu m/ms$. On the left, $a_{11}=-0.15 nm$, $a_{12}=-0.15 nm$.
On the right,  $a_{11}=-0.15 nm$, $a_{12}=-0.45 nm$; it can be observed how the second species is
extracted in the first excited state of the potential in Eq. (\ref{eqsolit}).
}
\label{fig:capture}
\end{figure}

The situation changes for small $g_{11}$, when the soliton size is comparable to
the trap size, Fig. \ref{fig:capture} --- a situation which may be difficult to achieve
experimentally but which we believe  is worth discussing. 
The dimensionless parameters of the simulations were $\eta_0=-50$ and $v=0.2$ for the two plots, taking $g_{11}=-0.5$, $g_{12}=-0.5$ on the left plot and $g_{11}=-0.5$, $g_{12}=-1.5$ on the right one. The dimensional parameters in the caption of the figure were calculated using the same considerations as
 for Fig. \ref{fig:reflect}. In Fig. \ref{fig:capture}, it turns out that the 
 soliton captures part of the cloud of the second species, which comes out of the
trap in one of the discrete eigenstates of (\ref{eqsolit}) (appropriately
Galileo-transformed to have the same velocity as the soliton).

In summary, there are three types of qualitative behaviour. The soliton can
traverse the trap leaving it undistorted (fig \ref{fig:reflect}a), it can push atoms out of the
trap without capturing them so the outcoupled atoms form a dispersive wave
(fig \ref{fig:reflect}b) or it can capture atoms of the second species so the outcome is
a vector soliton (fig \ref{fig:capture}).

\begin{figure}[htb]
{\centering  \resizebox*{0.46\columnwidth}{0.52\columnwidth}{\includegraphics{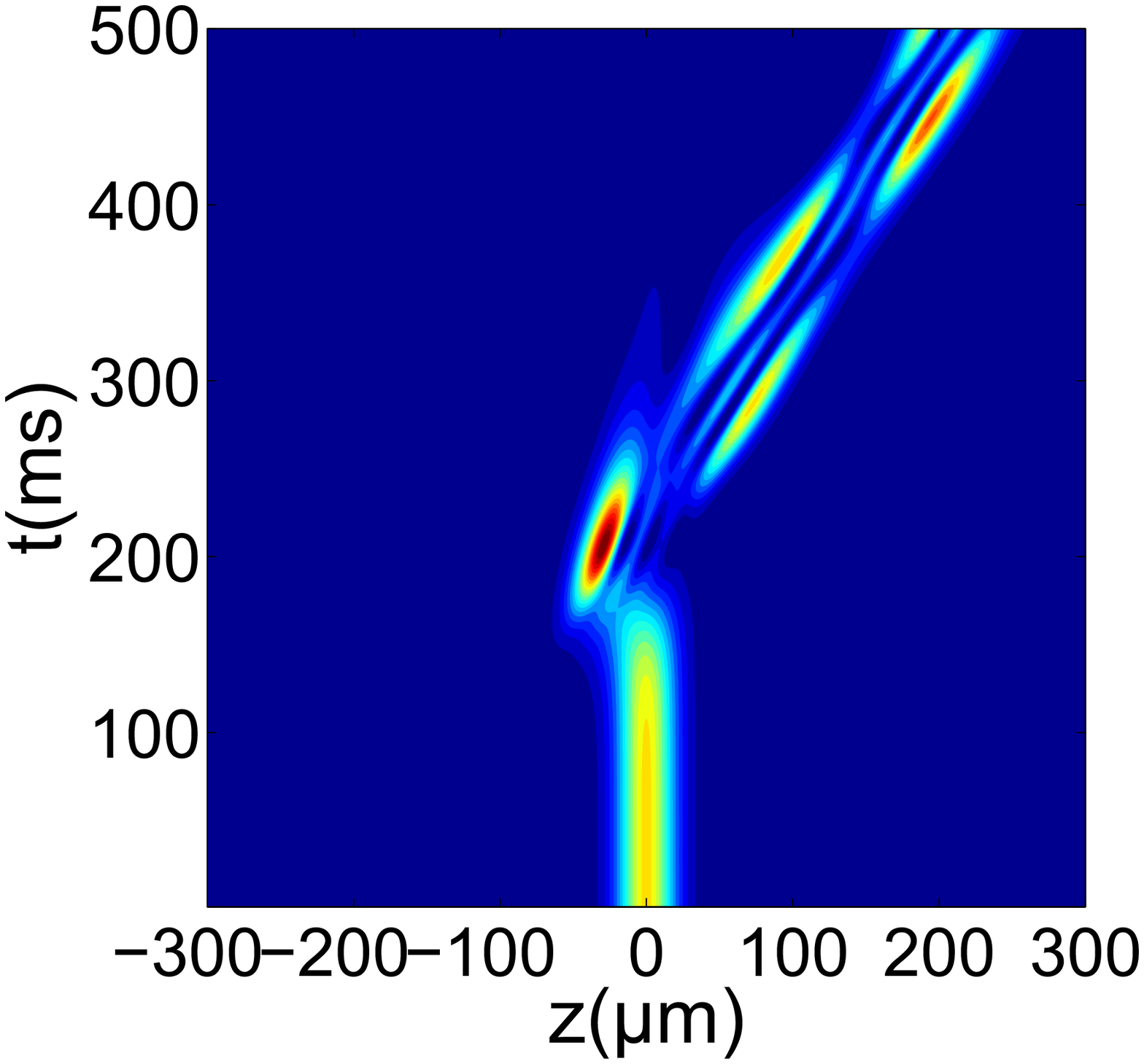}} \centering 
\resizebox*{0.52\columnwidth}{0.52\columnwidth}{\includegraphics{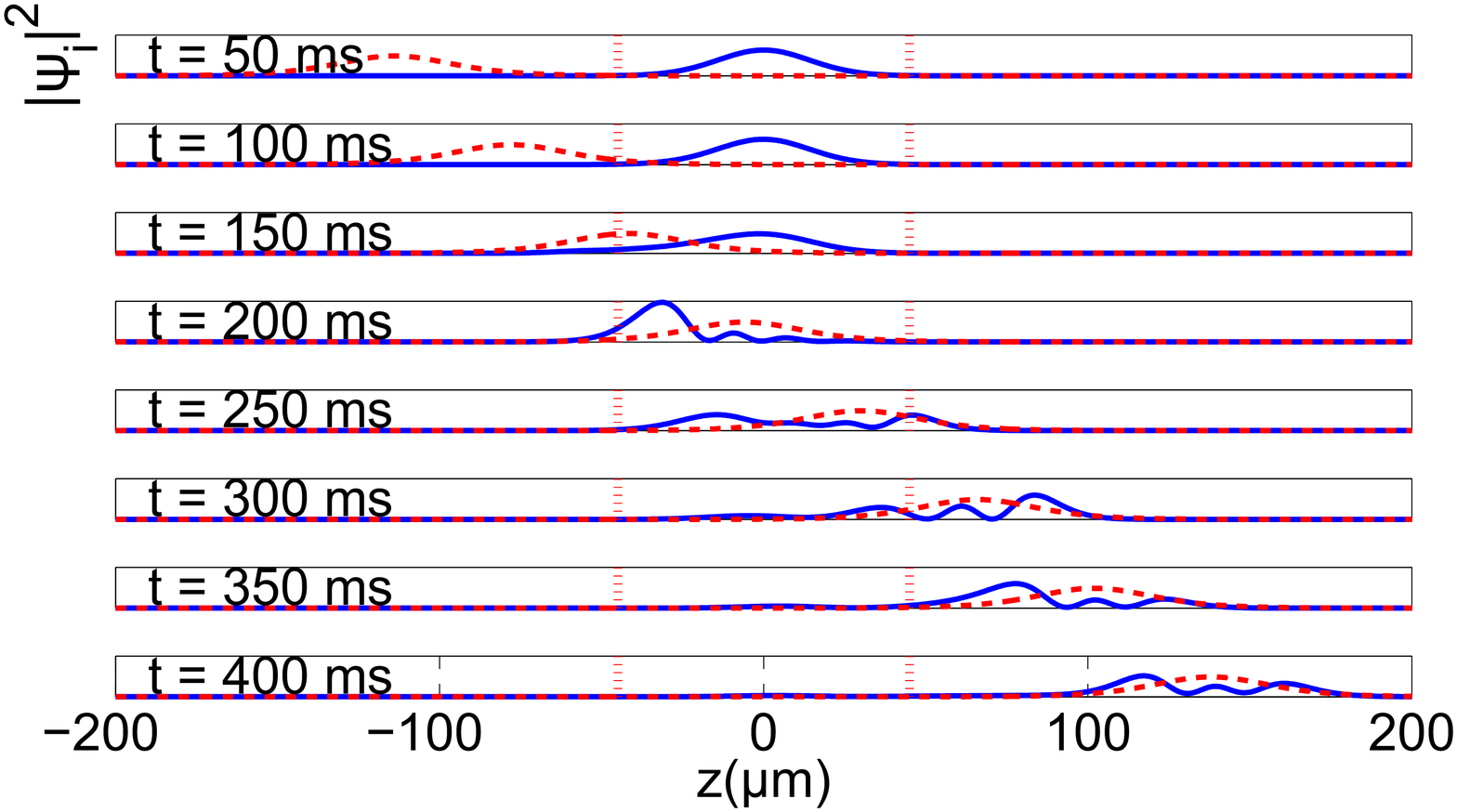}}\par}
\caption{(Color online). Almost complete extraction $a_{11}=-0.075 nm$, $a_{12}=-0.84 nm$,
$v=0.72\mu m/ms$. 
}
\label{fig:toy2}
\end{figure}

It is also interesting to discuss the dependence on $v$ of the number of extracted atoms. The precise behaviour depends on the parameters $g_{11}$, $g_{12}$ but, in general, the fraction of out-coupled atoms has a maximum at a certain velocity and then decreases for larger velocities, as it could be expected since reflection from a potential is typically smaller for larger momentum.
Nevertheless, for small values
of $|g_{11}|$ there are more complicated behaviours. The wave-function $|\psi_2|^2$ starts oscillating
around the moving soliton inside the trap. For some particular values of $v$, it coincides
that the soliton exits the trap while the oscillation of the cloud around it moves
rightwards. This sort of resonance enhances the atom out-coupling. An example is shown in Fig. \ref{fig:toy2}, where more than 90\% of the atoms are extracted from the trap. The simulation was performed with the dimensionless parameters $\eta_0=-50$, $v=0.2$, $g_{11}=-0.25$ and $g_{12}=-2.8$. The dimensionful parameters are presented again attending to the considerations exposed in Fig. \ref{fig:reflect}.

\begin{figure}[htb]
{\centering \resizebox*{0.8\columnwidth}{0.48\columnwidth}{\includegraphics{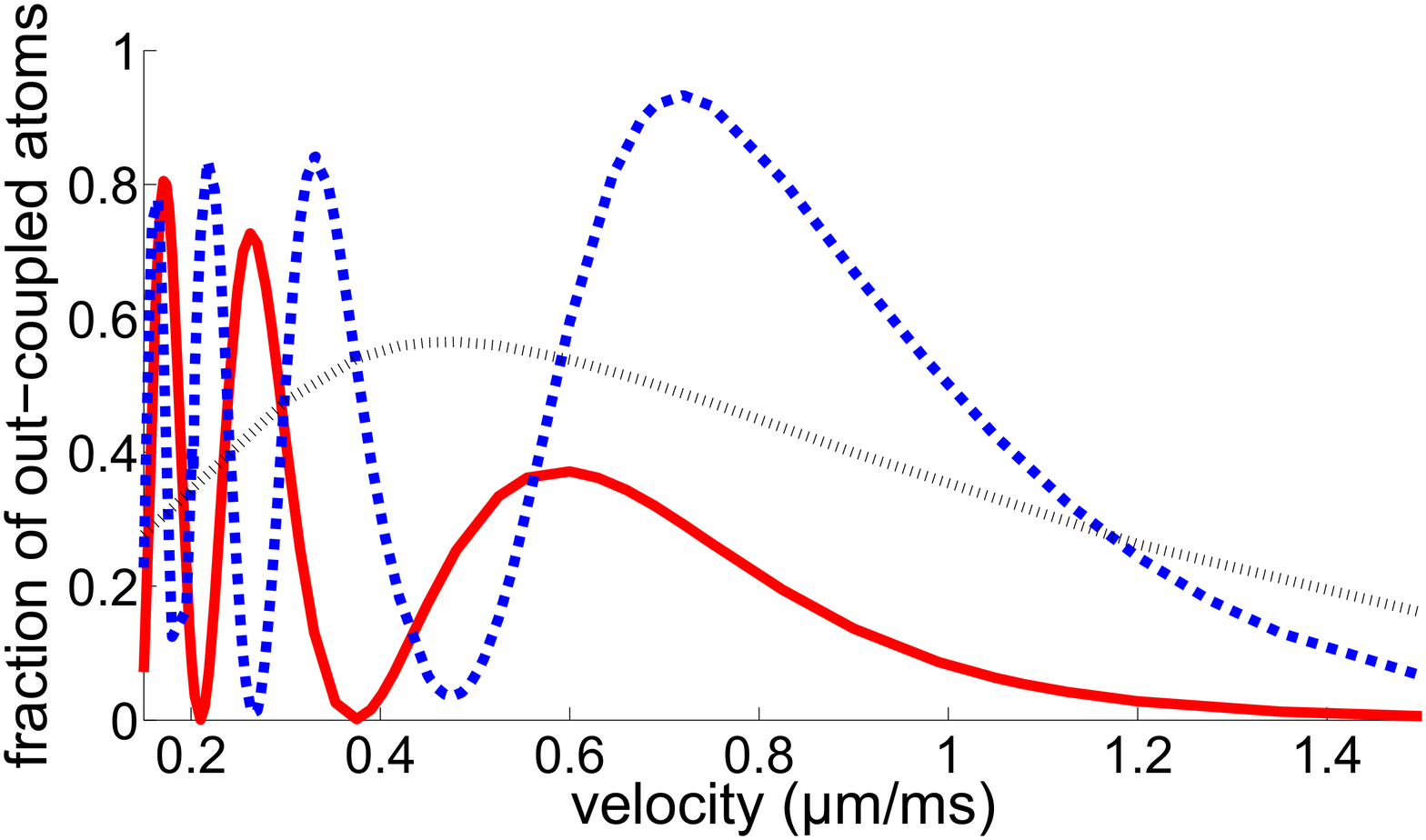}} 
\resizebox*{0.8\columnwidth}{0.48\columnwidth}{\includegraphics{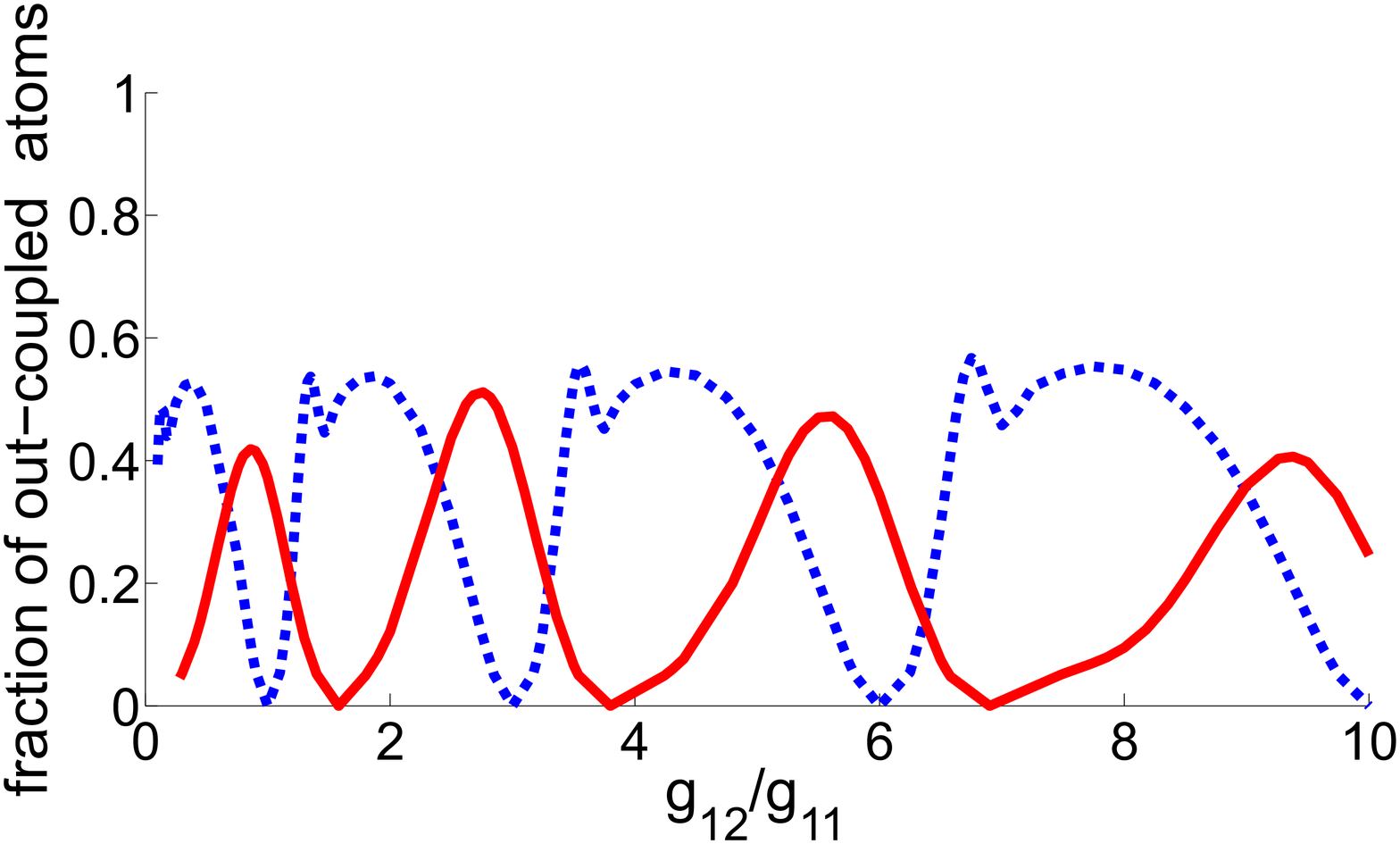}}\par}
\caption{(Color online). Fraction of out-coupled atoms. Top: as a function of 
velocity. The three graphs correspond to $g_{11}=-0.5$ ($a_{11}=-0.15 nm$), $g_{12}=-0.5$ ($a_{12}=-0.15 nm$) for the solid red line; 
$g_{11}=-0.25$ ($a_{11}=-0.075 nm$), $g_{12}=-2.8$ ($a_{12}=-0.84 nm$) for the dashed blue line;
 $g_{11}=-2$ ($a_{11}=-0.075 nm$), $g_{12}=-3.7$ ($a_{12}=-1.11 nm$) for the dotted black line.
Bottom: as a function of $g_{12}/g_{11}$. The two graphs correspond to $g_{11}=-2$ ($a_{11}=-0.6 nm$), $v=0.2$ ($v=0.6 \mu$m/ms) for the dashed blue line;
$g_{11}=-0.5$ ($a_{11}=-0.15 nm$), $v=0.2$ ($v=0.6 \mu m/ms$) for the solid red line.
}
\label{fig:data}
\end{figure}

In Fig. \ref{fig:data}, some data obtained from the numerical simulations
which condense the discussion of this section are presented.
The fraction of out-coupled
atoms are represented in terms of the different dimensionless parameters. Again, the dimensionful parameters are calculated as in the previous figures. 

%-----------------------------------------------

\subsection{B. Both species are initially trapped}

Now we start with the two species confined in the trap with the idea of controlling species 2 with the first one. The evolution of the first species is driven by modifying in time its correspondent intra-species parameter $a_{11}$ to produce solitons by modulational instability, along the lines of [13]. If $N_2 \ll N_1$, the presence of the other species affects this process only mildly. If the interspecies
force is attractive, these out-coupled solitons can capture matter from the second species, with the possibility of forming vector solitons. Indeed, with $a_{12}<0$, the first species generates "local traps" where the atoms of the second species tend to fall. 

It is also important that the initial Gaussian profile for the second atom cloud $w_{2,0}$ is wider than that of the first one $w_{1,0}$. The reason is that, when the solitons are formed in the first species, they cannot be very far of the atoms of the second species, or otherwise they would hardly trap them. Thus, for the $t<0$ situation, one needs $a_{22}>0$ such that $a_{22}>a_{11}$. For given $a_{ij}$ , both values of the Gaussian width profiles, $w_{1,0}$  and $w_{2,0}$ , can be determined minimising the functional $E_{TF}$ of section II, see Eq. (\ref{etf}).

\begin{figure}[htb]
\includegraphics[width=0.50\textwidth]{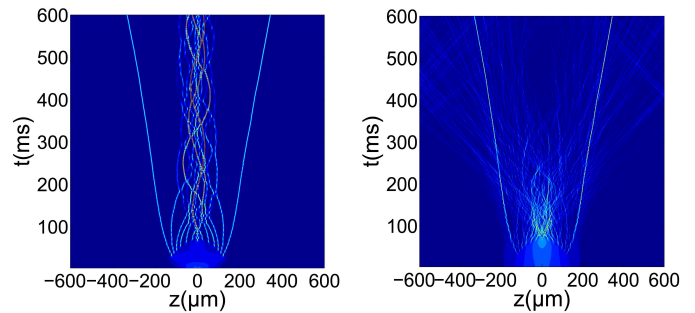}
\caption{(Color online). Emission of a pair of vector solitons with
$N_1 \gg N_2$. On the left, the plot representing lithium; on the right, the
rubidium. The parameters entering the simulation are presented in the text.}
\label{fig:extraction_LiRb}
\end{figure}

A concrete example is furnished by
considering a $^7$Li-$^{87}$Rb mixture in their lower hyperfine states. There are a broad intraspecies
resonance for lithium \cite{solitons1,7Li} and a broad interspecies resonance \cite{marzok} such that:
\begin{eqnarray}
a_{11}&=&-25 a_0 \left(1+\frac{192.3}{B-736.8}\right)\nonumber\\
a_{12}&=&-36 a_0 \left(1+\frac{70}{B-649}\right)
\end{eqnarray}
where $B$ is the external magnetic field in Gauss. For $^{87}$Rb, the background value
will be fixed $a_{22} = 100 a_0$ \cite{Fesh-review}.
The mass ratio is $\kappa=0.081$ and we take $N_1=5\times 10^4$,
$N_2 = N_1/10$. As in previous section, $\omega_{1\bot}=\omega_{2\bot}=10^{-3}$s 
such that $r_\bot=3\mu$m. The parameters of the trap will be taken to be 
$\tilde L=30$
(namely $L=90\mu$m) and $\tilde V_0 = 0.18$. Initially, the magnetic field is set
to be $B=579 $G ($a_{11}=0.29$nm, $a_{12}=0$) and the initial atom distributions have
$w_1 = \tilde L$, $w_2=1.73 \tilde L$. At $t=0$, the magnetic field is switched to
$B=697$G ($a_{11}=5.1$nm, $a_{12}=-4.7$nm) and the stretching of the cloud commences.
At $t=8$ms, the field is switched off $B=0$ ($a_{11}=-1$nm, $a_{12}=-1.7$nm) in order
to initiate soliton formation. The result is shown in figure \ref{fig:extraction_LiRb}.
As expected, the evolution of ${}^7$Li is essentially unaffected by the less numerous
rubidium atoms and the corresponding plot is similar to those found in \cite{paper1}.
On the right plot, it can be observed how a set of rubidium atoms are trapped by the
lithium consequently forming a vector soliton.
 It can also be appreciated that there is a number of atoms of this
  second species which are out-coupled from the trap which are not packed in the soliton.

\section{V. Creating supersolitons}

When solitons of two different species with repulsive interactions between them form an alternating array, 
they can behave as solid spheres in a Newton's cradle-like set-up. Thus, a perturbation propagates
along the array as an undistorted wave, analogous to a phonon, which was dubbed {\it supersoliton} in \cite{supersolitons}.
Modulational instability was mentioned in \cite{supersolitons} as the natural way of seeding such
an alternating array. The goal of this section is to explicitly show how this 
 can be accomplished by suitably manipulating in time the inter-atomic forces.

\begin{figure}[htb]
{\centering \resizebox*{\columnwidth}{0.5\columnwidth}
{\includegraphics{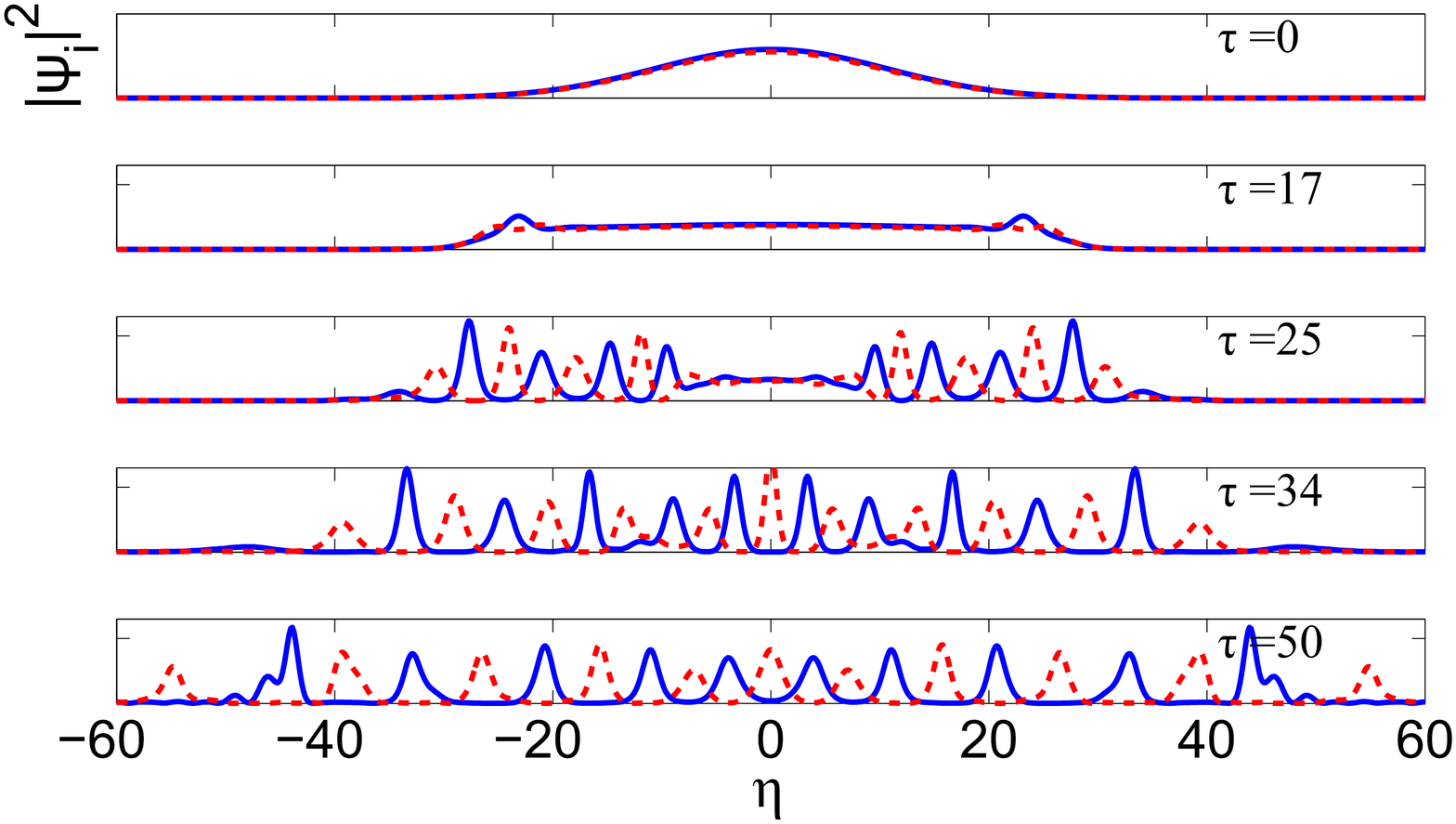}} 
\par}
\caption{(Color online). In this graph, $g = g_{12}=15$ for $\tau<\tau_s=10$ whereas 
$g=-20$ and $g_{12} =15$ for $\tau>\tau_s$. The rest of parameters are $\kappa=1$ , $N_2=0.95 N_1$
}
\label{fig:super}
\end{figure}

For concreteness, we fix $\kappa=1$ and $g_{11}=g_{22}=g$,  but it should be understood that 
these conditions are not essential for the qualitative features of the process. 
A slight asymmetry is introduced $N_2=0.95 N_1$ in order to avoid that $\psi_1=\psi_2$ is a solution,
see section III. In any case, the supersoliton excitation could also be produced in a fully symmetric
set-up since in this case the symmetric solution tends to be unstable during the whole process:
in a situation with 
$g_{11} < 0$, $g_{22} < 0$, $g_{12} > 0$, modulational instability always sets in
\cite{kasamatsu,vector_modul}.
The following protocol is considered: initially, all scattering lengths are repulsive,
 producing the stretching of the atom cloud.
Then, the intraespecies scattering lengths are tuned to negative values producing the instability.
Due to interspecies repulsion, the system can evolve into the desired alternating array
of solitons, see Figure \ref{fig:super}. 

%is an example of the process, the considered dimensionless parameters where $g=15$, $g_{12}=15$  for $\tau<\tau_s=10$ whereas 
%$g=-20$ and $g_{12} =15$ for $\tau>\tau_s$. %These quantities were introduced fixing $\omega_{1\bot}=\omega_{2\bot}=10^3$s$^{-1}$ and taking $r_\bot \approx -\mu$ (mixture --), so they correspond to the values of the dimensional parameters displayed in the caption of the figure.

\begin{figure}[htb]
\includegraphics[width=0.48\textwidth]{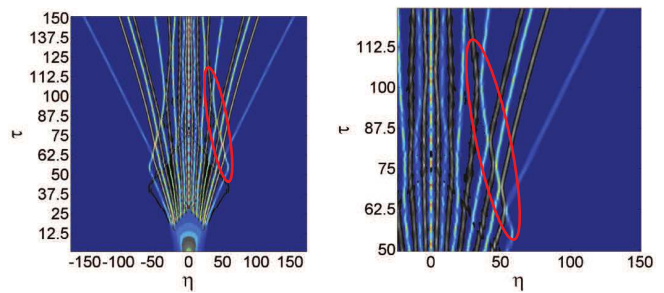}
\caption{(Color online). The left plot shows how the system acts like
an array of hard balls colliding among themselves. 
On the right, the region of the supersoliton-like excitation is enlarged and indicated with a red ellipse.}
\label{fig:super2}
\end{figure}

Once the alternating array is formed and in order to produce a kick,
 a potential wall can be placed in the way of the fastest soliton. The soliton bounces back to
collide with the adjacent one therefore starting a supersoliton-like excitation in the sample,
providing a simple realization of the ideas of \cite{supersolitons}.
 In particular, it is illustrative to implement
a time-dependent step-like potential:
\begin{equation}
f_{s} = f_{s,0}(\tanh(\eta-\eta_s) - \tanh(\eta+\eta_s) 
+2\tanh(\eta_s))\Theta(\tau_{b}-\tau) 
\label{wall}
\end{equation}
where $\Theta(x)$ is Heaviside function and
$\tau_{b}$ is the time of the bounce of the fastest soliton at the step.
The expression in Eq. (\ref{wall}) has to be added to the Gaussian trap (\ref{trap}).
Figure \ref{fig:super2} shows how the solitons act as hard
balls because of their strong inter-repulsion.

The number of atoms per soliton can be easily estimated by integrating $|\psi_i|^2$ in
the relevant region. As expected, once the alternating array is formed, the number of
atoms per soliton does not change upon latter time evolution.  This can be also checked
in a non-symmetric case with $\kappa \neq 1$, where
the qualitative behaviour is analogous to the above described.

%after taking a mixture of --- because of ---, for the last figure $\kappa=1$ and $g_{11}=g_{12}$, $N_2=0.90 N_1$, $g_{11} = g_{12}=15$, $g_{22}=-1$  for $\tau<\tau_s=10$ whereas $g_{11}=-20$, $g_{22}=-1$ and $g_{12} =15$ for $\tau>\tau_s$.

%%%%%%%%%%%%%%%%%%%%%%%%%%%%%%%%%%%%%%%%                                 

\subsection{A. Experimental feasibility}

The described process requires a simultaneous tuning of both intra-species scattering
lengths while the inter-species one remains fixed. For magnetically tuned Feshbach
resonances --- see Eq. (\ref{aofB}) --- approximately equal values of $B_0$ would be needed for both
species. We are not aware of any mixture in which this coincidence happens and, in fact, it
can be considered a marginal situation. An intriguing possibility would be to use optical
methods \cite{opticalFB} together with magnetic ones in order to independently displace the
two closed channel levels and make the Feshbach resonances coincide. However, since we are not
aware of any experimental result in this direction, 
we leave the previous discussion as a theoretical possibility
and now look for situations
in which similar processes can be achieved by tuning just one of the three relevant 
scattering lengths.

With this aim, let us consider a BEC composed by ${}^{7}$Li  and 
${}^{23}$Na, both in the hyperfine
$|F=1\ m_f=1\rangle$ state. One can make use of the intra-species resonance around
$B=736.8$ G for ${}^{7}$Li \cite{7Li} which allows to tune the value of $a_{11}$. 
The other intra-species and the
inter-species interaction would correspond to their
background values $a_{22} \approx 63 a_0$
\cite{Fesh-review} and $a_{12}\approx 20 a_0$ \cite{schuster}.

\begin{figure}[htb]
\includegraphics[width=0.48\textwidth]{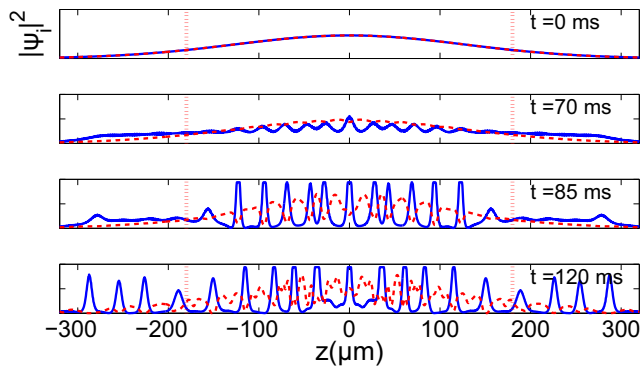}
\caption{(Color online). In this graph, $a_{11}=7.75 nm$, $a_{22}=63 a_0$ and $a_{12}=20 a_0$ for $t<t_s=40 ms$ whereas 
$a_{11}=-1.75 nm$, $a_{22}=63 a_0$  and $a_{12}=20 a_0$ for $t>t_s$. The rest of parameters are $\kappa=0.3061$, $L=180 \mu m$, $V_0=0.75 \hbar \omega_\perp$, $N_1=N_2=30000$.
}
\label{fig:supersoliton3}
\end{figure}

Figure \ref{fig:supersoliton3} shows the formation of the array for this case. 
It can be appreciated how  species 1 forms solitons and the species 2 dispersive waves,
since $a_{22}$ is kept positive. 
Fixed $g_{22}=67$ and $g_{12}=22$, the taken dimensionless scattering lengths 
were $g_{11}=155$ for $\tau<\tau_s=40$ whereas 
$g_{11}=-35$ for $\tau>\tau_s$. The rest of parameters were $\kappa=0.3061$, $\tilde L=60$, $\tilde V_0=0.75$ and $N_1=N_2=30000$. The correspondent dimensionful values are displayed in the caption of the figure.

\begin{figure}[htb]
\includegraphics[width=0.48\textwidth]{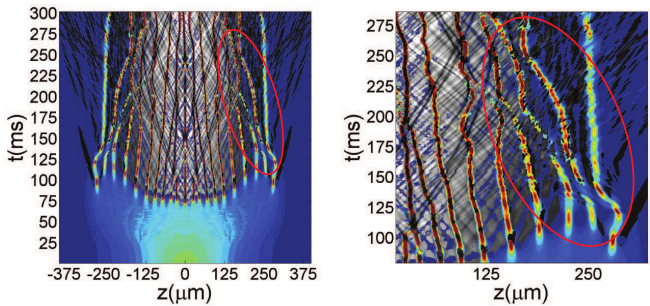}
\caption{(Color online). Example of pseudo-super-solitons colliding after the formation of the array in a mixture of ${}^7$Li-{}$^{23}$Na. On the right, the region of the supersoliton-like excitation, which is indicated with a red ellipse, is enlarged.}
\label{fig:supersoliton4}
\end{figure}

Again, the potential wall of Eq. (\ref{wall}) is added to the Gaussian trap (\ref{trap}). In  figure \ref{fig:supersoliton4} the soliton of species 1 bounces back to collide
with the matter waves of species 2, therefore sparking a pseudo-supersoliton
excitation.
Notice that species 2 plays a crucial role in the process even if its atoms are not
packed in solitons themselves. 

%This resonance has been crucial in the attainment of a BEC with only ${}^{7}$Li
%atoms, in the formation of bright soliton after its collapse
%and in the achievement of the mentioned two-species condensate.

Another mixture for which one can conceive a similar phenomenon is
that composed by ${}^{85}$Rb in its hyperfine
$|F=2\ m_f=-2\rangle$ state and 
${}^{87}$Rb with $|F=1\ m_f=-1\rangle$ \cite{RbRb}. Near the intra-species resonance around
$B=155$ G for ${}^{85}$Rb \cite{claussen}, it is possible to tune the value of $a_{11}$. 
This resonance has been crucial in the attainment of a BEC with only ${}^{85}$Rb
atoms \cite{cornish}, in the formation of bright soliton after its collapse
\cite{cornish2} and in the achievement of the mentioned two-species condensate
\cite{RbRb}. Around this value of the magnetic field, the other intra-species and the
inter-species interaction are away from any resonance and correspond to their
background values $a_{22} \approx 100 a_0$ \cite{mertes}
and $a_{12}\approx 213 a_0$ \cite{papp}. Nevertheless, an eventual experiment with
this mixture may be more delicate because of the large value of $a_{12}$, which can
induce  modulational instability in the initial phase
 when all scattering lengths remain positive.

\section{VI. Discussion}

The degree of control of cold atomic clouds that has been experimentally achieved
is remarkable. In this contribution, we have theoretically discussed several possibilities of
atom manipulation of two-component BECs trapped in quasi-one-dimensional potentials, by
tuning in time the different inter-atomic forces. 
We have shown that there are several interesting phenomena that can occur in multicomponent
systems and that do not have any obvious counterpart in the single species case. 
Having both inter- and intra-species scattering which can be separately tuned in
time and/or space opens a plethora
of possibilities of which we have explored a few.
We emphasize that numerical simulations indicate that 
the qualitative behaviours shown are rather robust in the sense
that they do not depend crucially on the shape of the trap, the initial profile or
the particular atomic species for which the computations are done. They do depend,
though, on the concrete values of the $a_{ij}$ and the instants chosen for their
tuning. Inappropriate values can result in a failure in the formation or the out-coupling
of solitons. 
In any case, the essential point
is the ability to externally control the interatomic forces in the desired ways.

Markedly, 
there are a number of different protocols that can be used in order to out-couple matter from
shallow traps, providing atom laser-like devices. It can be of particular interest the
possibility of using matter to manipulate matter as discussed in section IV since this could
eventually
lead to a sort of {\it atomic tweezers} that might generalize the laser tweezers which 
are readily used for atom manipulation, see for instance \cite{tweez}  for recent
interesting work.

In principle, the processes analyzed in this paper could be utilized as a 
kind of coherent beam splitter, the first step in a protocol of atom interferometry. It would
be very interesting to understand the possibilities and limitations that these systems could
have in terms of precision metrology, but this lies beyond the scope of the present work.

Since the whole analysis is based on solving the one-dimensional Gross-Pitaevskii equations, it
may be possible that the results presented could be relevant for other physical systems apart from
Bose-Einstein condensates, as for example light propagation in optical fibers --- see
for instance
\cite{saleh} for a recent work which shares some similarities with the set-up of section IV-A.

\

\centerline{{\bf Acknowledgements}}

%-------------------------------------------------------------------------
We thank David N\'ovoa for helpful discussions.
This work was supported by Xunta de Galicia (project 10PXIB383191PR) and by the
Univ. Vigo research programme.
The work of A. Paredes is supported by the Ram\'on y Cajal programme. 
The work of D. Feijoo is supported by the FPU Ph. D. fellowship programme.

%--------------------------------------------------------------------------

\end{document}